\documentclass[11pt, letter,fleqn]{article}
\usepackage{amsmath, amssymb,amsthm,cite}
\usepackage{chngcntr}
\usepackage{url}

\title{Symbolic Computation of Equivalence Transformations and Parameter Reduction for Nonlinear Physical Models}

\author{
Alexei F. Cheviakov\footnotemark[2], \\
{\small \emph{Department of Mathematics and Statistics, University of Saskatchewan, Saskatoon, S7N 5E6 Canada}}\\[3ex]
}

\newtheorem{theorem}{Theorem}

{\theoremstyle{definition} \newtheorem{remark}[theorem]{Remark}}
{\theoremstyle{definition} \newtheorem{example}[theorem]{Example}}
{\theoremstyle{definition} \newtheorem{definition}[theorem]{Definition}}

\def \dfracskip{2ex}

\def\eps{\varepsilon}

\def\const{\hbox{\rm const}}

\def\PDEs#1#2#3{{\boldsymbol{\rm #1}}\{#2\,; #3\} }
\def\sg#1{{\rm #1}}
\def\beq{\begin{equation}}
\def\eeq{\end{equation}}
\def\barr{\begin{array}{ll}}
\def\earr{\end{array}}

\begin{document}
\footnotetext[2]{Corresponding author. Electronic mail: chevaikov@math.usask.ca}

\counterwithin{equation}{section}

\maketitle

\begin{abstract}

An efficient systematic procedure is provided for symbolic computation of Lie groups of equivalence transformations and generalized equivalence transformations of systems of
differential equations that contain arbitrary elements (arbitrary functions and/or arbitrary constant parameters), using the software package \verb|GeM| for \verb|Maple|. Application of equivalence transformations to the reduction of the number of arbitrary elements in a given system of equations is discussed, and several examples are considered. First computational example of a generalized equivalence transformation where the transformation of the dependent variable involves the arbitrary constitutive function is presented.

As a detailed physical example, a three-parameter family of nonlinear wave equations describing finite anti-plane shear displacements of an incompressible hyperelasic fiber-reinforced medium is considered. Equivalence transformations are computed and employed to radically simplify the model for an arbitrary fiber direction, invertibly reducing the model to a simple form that corresponds to a special fiber direction, and involves no arbitrary elements.

The presented computation algorithm is applicable to wide classes of systems of differential equations containing arbitrary elements.

\end{abstract}

\section{Introduction}

The  majority of contemporary mathematical models, in particular, physical models written in terms of partial or ordinary differential equations (PDEs, ODEs), involve constant or variable parameters. These parameters, also called arbitrary elements, appear in the equations in the form of arbitrary constants that assume values in certain ranges, and/or arbitrary functions belonging to  certain classes. Equivalent formulations of a model at hand, involving fewer constitutive parameters and/or reduced forms of constitutive functions, can lead to a significant simplification of the analysis of the model, especially when such analysis involves classifications. Such equivalent formulations can be systematically sought using \emph{equivalence transformations}.

Equivalence transformations of a family of DE (differential equation) systems are transformations of problem variables and arbitrary elements that map every DE system from the given family into another system from the same family. As a basic illustration, consider a set of (1+1)-dimensional nonlinear diffusion equations for $u(x,t)$, given by
\beq\label{eq:1}
u_{t}=c^2(u) u_{xx},
\eeq
where $c(u)$ is an arbitrary constitutive function. (In \eqref{eq:1} and throughout the paper where appropriate, subscripts denote derivatives.) The family \eqref{eq:1} admits, for example, an obvious discrete equivalence transformation $c^*(u)=-c(u)$; in particular, any solution $u(x,t)$ of \eqref{eq:1} is mapped a solution of a PDE $u_{tt}=(c^*(u))^2 u_{xx}$. Scalings and translations are the simplest kinds of continuous equivalence transformations. In particular, transformations that re-scale variables using their typical values, and thus map a problem into a dimensionless one, or scaling transformations in general, have a long history, and are commonly used to simplify and non-dimensionalize DE problems. For example, the PDE family \eqref{eq:1} clearly admits scaling and translation-type equivalence transformations
\beq\label{eq:1:sc}
x^* = A_3 x + A_1,\quad t^* = A_4 t + A_2,\quad u^*(x^*,t^*) = A_5 u(x,t),\quad  c^*(u^*) =  \dfrac{A_3^2}{A_4} c(u),
\eeq
$A_1, \ldots, A_5\in \mathbb{R}$, $A_4>0$, so that a solution $u(x,t)$ of a PDE \eqref{eq:1} with a constitutive function $c(u)$ is mapped into a solution $u^*(x^*,t^*)$ of a PDE
\[
u^*_{t^*}=(c^*(u^*))^2 u^*_{x^*x^*},
\]
a member of the family \eqref{eq:1} with a  different 
constitutive function, as long as $A_3^2\ne A_4$.

The notion of equivalence transformations is closely related to that of \emph{symmetries} of a system of differential equations. A symmetry is a transformation that leaves invariant the solution manifold of a given system, i.e., maps any solution of the system into a solution of the same system. Due to the general topological nature of this definition, a systematic way to calculate \emph{all} symmetries of a given DE system, and similarly, \emph{all} equivalence transformations of a given family of DE systems, is not available. However, equivalence transformations of particular kinds, such as Lie groups of local and nonlocal equivalence transformations, can indeed be computed systematically.

Work on equivalence transformations was initiated in \cite{ovsiannikov2014group}. Various examples, extensions, and applications of the notion of equivalence transformations appear in the literature, such as \cite{BCABook,lisle1992equivalence,meleshko1996generalization, ibragimov1991preliminary,bila2011new,vaneeva2009enhanced, bluman2007nonlocally} and references therein. An excellent review and work on equivalence transformations, including both theoretical and computational aspects, is contained in the thesis \cite{lisle1992equivalence} (see also \cite{Huang2009}).


The current manuscript presents an algorithmic procedure to calculate Lie groups of local equivalence transformations of a given DE system using a symbolic symmetry and conservation law computation package \verb"GeM" for \verb"Maple" \cite{cheviakov2007gem,cheviakov2010computation,cheviakov2010symbolic} developed by the author. A brief review of the notion of equivalence transformations, relations between Lie groups of point equivalence transformations and Lie point symmetries, and extended classes of equivalence transformations, is given in Section \ref{sec:equiv:theor}. The Section also contains several illustrative examples, and ends with an outline of a systematic procedure of computation of Lie groups of generalized equivalence transformations, closely related to the algorithm of point symmetry computation. Section \ref{sec:algo:gem} contains a typical sequence of steps for complete symbolic computation of Lie groups of generalized equivalence transformations using the \verb|GeM|  \verb|Maple| package.

Section \ref{sec:eqtr:egs} is devoted to three examples of symbolic computation of equivalence transformations of PDE models. Complete detailed \verb|Maple| code is included for each example. In the first example, equivalence transformations are computed for the well-known three-parameter family of dimensional Korteweg-de Vries (KdV) equations, and are used to invertibly reduce the family to a single canonical KdV equation involving no arbitrary parameters. The second example is devoted to the calculation of nonlocal equivalence transformations of a family of nonlinear wave equations. In particular, nonlocal equivalence transformations are computed as point transformations of a potential system, and do not correspond to local equivalence transformations of the model. The final example in Section \ref{sec:eqtr:egs} is a Lie group of generalized equivalence transformations of a PDE system, where the transformation component corresponding to the dependent variable involves an explicit dependance on an arbitrary element that is an arbitrary function. This is the first computational example of this situation in the literature (see Remark \ref{rem:mel:false}).

Another example considered in Section \ref{sec:eqtr:1fiber} presents a recent model of finite anti-plane shear displacements in nonlinear incompressible hyperelasic fiber-reinforced hyperelastic materials \cite{cheviakov2015fully}. Such models arise in the description of multiple types of  nonlinear media, including biological membranes \cite{cheviakov2015one}. The systematically computed local equivalence transformations are used to drastically simplify the model, by invertible transformations, through the complete elimination of arbitrary elements. Together with the classical example of the Korteweg-de Vries model in Section \ref{sec:eqtr:egs}, the example of Section \ref{sec:eqtr:1fiber} addresses the questions of primary importance for the analysis of nonlinear equations arising in physics and applied science -- the identification of essential parameters of the model, and the possibility of invertible parameter reduction.

Section \ref{sec:concl} concludes the paper with a discussion of the presented algorithms, their efficiency, results, related classification problems, and an overview of open questions and possible work directions in the area.

Throughout the paper, summation in repeated indices is assumed where appropriate.


%
%
%
%
%
%

%

\section{Equivalence Transformations, their Computation, and Extensions}\label{sec:equiv:theor}

\subsection{Lie Groups of Equivalence Transformations}\label{sec:equiv:theor:LG}

Consider a family $\mathcal{F}_K$ of DE systems $\PDEs{R}{x}{u; K}$:
\beq\label{eq:ch1:sym:PDEsys_eqivtr}
R^\sigma(x,u,\partial u, \ldots , \partial^k u, K)=0, \quad \sigma=1,\ldots ,N,
\eeq
involving $n$ independent variables $x=(x^1,\ldots ,x^n)$, $m$ dependent variables
$u(x)=(u^1(x),\ldots ,u^m(x))$, and $L$ arbitrary elements (constitutive functions and/or parameters) $K=(K^1,\ldots,\,K^L)$.
In \eqref{eq:ch1:sym:PDEsys_eqivtr}, partial derivatives are denoted by $u^\mu_i={\partial u^\mu(x)}/{\partial
x^i}$. The symbol
\[
\barr
\partial^p u&=\left\{ u^\mu_{{i_1} \ldots {i_p}}~|~~\mu=1,\ldots ,m;~~i_1,\ldots ,i_p = 1,\ldots ,n \right\}\\[\dfracskip]
&=\left\{ \dfrac{\partial^p u^\mu(x)}{\partial x^{i_1}\ldots \partial x^{i_p}}~|~\mu=1,\ldots ,m; i_1,\ldots ,i_p = 1,\ldots ,n
\right\}
\earr
\]
is used to denote the set of all partial derivatives of order $p$, $p=1,2,\ldots$. The arbitrary elements $K$ in \eqref{eq:ch1:sym:PDEsys_eqivtr} may be either arbitrary constants or arbitrary functions within some class. Constitutive functions are assumed to be sufficiently smooth functions of dependent and/or independent variables and/or derivatives of the dependent variables, or combinations thereof.

In general, equivalence transformations for a DE family $\mathcal{F}_K$ \eqref{eq:ch1:sym:PDEsys_eqivtr} can be defined as follows.
\begin{definition} \label{def:equivalence:GenLisle}
An \emph{equivalence transformation} for a DE family $\mathcal{F}_K$ is a change of variables and arbitrary elements $(x,u,K)\to(x^*,u^*,K^*)$ which maps every DE system $\PDEs{R}{x}{u; K}\in \mathcal{F}_K$ into a DE system $\PDEs{R}{x^*}{u^*; K^*}\in \mathcal{F}_K$.
\end{definition}


Definition \ref{def:equivalence:GenLisle} is rather general, and there is no systematic way to compute explicit forms of transformations described therein. For the practical purposes, the following definition describing a special subclass of equivalence transformations is often used \cite{ovsiannikov2014group, lisle1992equivalence}.
\begin{definition} \label{def:equivalence:std}
A \emph{one-parameter Lie group of point equivalence transformations} of a family $\mathcal{F}_K$ of DE systems \eqref{eq:ch1:sym:PDEsys_eqivtr} is a
one-parameter Lie group of transformations acting in the space of $(x,u,K)$, given by
\beq \label{eq:ch1:symm:equivalence:eq_trans}
\begin{array}{llll}
(x^*)^i&=&f^i(x,u;\eps)=x^i + \eps \xi^i(x,u) + O(\eps^2),\quad i=1,\ldots ,n,\\
(u^*)^\mu&=&g^\mu(x,u;\eps)=u^\mu + \eps \eta^\mu(x,u) + O(\eps^2),\quad \mu=1,\ldots ,m,\\
(K^*)^\ell&=&G^\ell(Q^\ell;\eps)={K}^\ell + \eps \kappa^\ell(Q) + O(\eps^2),\quad \ell=1,\ldots, L,
\end{array}
\eeq
which maps a PDE system $\PDEs{R}{x}{u; K} \in \mathcal{F}_K$ into another PDE system
$\PDEs{R}{x^*}{u^*; K^*}$ in the same family.
\end{definition}
In formulas \eqref{eq:ch1:symm:equivalence:eq_trans}, the arguments $Q^\ell$ of the corresponding transformation $G^\ell$ and the tangent vector field component $\kappa^\ell$ depend on the nature of the arbitrary element $K^\ell$, $\ell=1,\ldots, L$. In particular, if $K^\ell$ is a constant parameter, $Q^\ell$ may be the set of all constant parameters of the DE system family \eqref{eq:ch1:sym:PDEsys_eqivtr}. If $K^\ell$ is a arbitrary (constitutive) function, $Q^\ell$ may also involve independent/dependent variables $K^\ell$ depends on, other constitutive functions with compatible dependencies, etc. Within Definition \ref{def:equivalence:std}, transformations \eqref{eq:ch1:symm:equivalence:eq_trans} form a group on the $(x, u)$ space; the transformations for $x, u$ do not involve arbitrary elements.


\begin{remark}
The notion of a Lie group of equivalence transformations \eqref{eq:ch1:symm:equivalence:eq_trans} of a DE system is intimately related to that of a Lie group of point symmetries; the appropriate notation, including that for infinitesimal generators, and further details, can be found in the following Section \ref{sec:eqtr:ps}.
\end{remark}

Multiple examples of equivalence transformations are known in the literature that go beyond the Definition \ref{def:equivalence:std}. Some of such examples are considered in Section \ref{subsec:eq:tr:gen} below.


\begin{example} The transformations \eqref{eq:1:sc} involve five one-parameter groups of equivalence transformations. Indeed, they can be written in the global group form as
\beq\label{eq:1:sc2}
x^* = e^{\eps_3} x+\eps_1,\quad t^* = e^{\eps_4} t+\eps_2,\quad u^*(x^*,t^*) = e^{\eps_5} u(x,t),\quad  c^*(u^*) =  e^{\eps_3-\eps_4/2} c(u),
\eeq
with group parameters $\eps_i$, $i=1,\ldots,5$. The corresponding infinitesimal generators are given by
\beq\label{eq:1:sc2X}
\sg{X}_1=\frac{\partial }{\partial x},\quad \sg{X}_2=\frac{\partial }{\partial t},\quad \sg{X}_3=x\frac{\partial }{\partial x}+c\frac{\partial }{\partial c},\quad \sg{X}_4=t\frac{\partial }{\partial t}-\frac{1}{2}c\frac{\partial }{\partial c}, \quad \sg{X}_5=u\frac{\partial }{\partial u}.
\eeq

%
\end{example}

\begin{example}[Equivalence transformations and parameter reduction]\label{eg:2:2}
As a second example, consider a classical family of  ``dimensional Korteweg-de Vries" equations for the unknown function $u(x,t)$, given by
\beq\label{eq:dimKdV:fam}
u_t + a u_x + buu_x +q u_{xxx}=0,
\eeq
involving three constant parameters $a,b,q\in \mathbb{R}$, $b,q\neq 0$. The model \eqref{eq:dimKdV:fam} arises in multiple  physical contexts \cite{su1969korteweg}, for example, it describes the evolution of the fluid surface elevation $u(x,t)$ in a limit considering small-amplitude long waves in a shallow channel. We now recollect the well-known equivalence transformations of the PDE family \eqref{eq:dimKdV:fam}. (An example of symbolic software computations of these equivalence transformations is given below in Section \ref{sec:eqtr:KdV}.)

A basic set of equivalence transformations of the PDE family \eqref{eq:dimKdV:fam}  is given by infinitesimal generators
\beq\label{eq:1:sc2X}
\barr
\sg{Y}_1=\dfrac{\partial }{\partial x},\quad \sg{Y}_2=\dfrac{\partial }{\partial t},\quad \sg{Y}_3=\dfrac{\partial }{\partial u}- b \dfrac{\partial }{\partial a},\quad
\sg{Y}_4=t\dfrac{\partial }{\partial x}+ \dfrac{\partial }{\partial a},\\[3ex]

\sg{Y}_5=x\dfrac{\partial }{\partial x} + t\dfrac{\partial }{\partial t}+ 2q\dfrac{\partial }{\partial q},\quad
\sg{Y}_6=x\dfrac{\partial }{\partial x} + 3t\dfrac{\partial }{\partial t}- 2u\dfrac{\partial }{\partial u}- 2a\dfrac{\partial }{\partial a},\\[3ex]
\sg{Y}_7=x\dfrac{\partial }{\partial x} + 3t\dfrac{\partial }{\partial t}- 2a\dfrac{\partial }{\partial a}- 2b\dfrac{\partial }{\partial b}.
\earr
\eeq
The corresponding point transformations can be written, for example, as
\beq\label{eq:KdV:eqivtr}
\barr
x^* = \dfrac{A_5}{A_6 A_7} (x-A_4 t) + A_1,\qquad t^* = \dfrac{A_5}{A_6^3 A_7^3} t + A_2,\qquad u^*(x^*,t^*) = A_6^2 \,(u(x,t)+A_3),\\[3ex]
a^*=A_6^2 A_7^2\,(a- A_3 b-A_4 ),\qquad b^*=A_7^2\,b,\qquad q^*=A_5^2\,q.
\earr
\eeq
with seven arbitrary constants $A_1, \ldots, A_7\in \mathbb{R}$.

The equivalence transformations \eqref{eq:KdV:eqivtr} can be used to invertibly map the PDE \eqref{eq:dimKdV:fam} into one of the standard forms of the KdV equation, for example,
\beq\label{eq:KdV}
u_t + uu_x + u_{xxx}=0,
\eeq
thus reducing the number of constant parameters from three to zero. To do this, one removes the term $a u_x$ ($a\ne 0$), and re-scales the variables. In common KdV derivations, one of two common ways is used to remove the term $a u_x$. This can be  achieved by passing to a traveling wave coordinate $x^*=x-at$, $t^*=t$, $u^*(x^*,t^*)=u(x,t)$, that is, using the transformation \eqref{eq:KdV:eqivtr} with
\[
A_1=A_2=A_3=0, \quad A_4=a, \quad A_5=A_6=A_7=1.
\]
Another way consists in a pure shift of the dependent variable, i.e., the transformation \eqref{eq:KdV:eqivtr} with
\beq\label{eq:dimKdV:trans:KillUx:2}
A_1=A_2=A_4=0, \quad A_3=a/b, \quad A_5=A_6=A_7=1,
\eeq
leading to the choice of variables $x^*=x$, $t^*=t$, $u^*(x^*,t^*)=u(x,t)+A_3$. Both of the above choices, as well as any other combined transformation leading to $a^*\propto a- A_3 b-A_4=0$ in \eqref{eq:KdV:eqivtr}, yield an equivalent PDE from the family \eqref{eq:dimKdV:fam}, given by (after the omission of the asterisks)
\beq\label{eq:dimKdV:fam:no:a}
u_t + buu_x +q u_{xxx}=0.
\eeq
Due to the freedom in changing the signs of $x$ and $u$, i.e., in the choice $A_5$, $A_6$, $A_7=\pm 1$ in \eqref{eq:KdV:eqivtr}, one can assume without loss of generality that in \eqref{eq:dimKdV:fam:no:a}, $b,q>0$. Finally, using again the transformations \eqref{eq:KdV:eqivtr} with
\[
A_1=A_2=A_3=A_4=0, \quad A_5=q^{-1/2},\quad A_6=1,\quad A_7=b^{-1/2},
\]
one transforms the PDE \eqref{eq:dimKdV:fam:no:a} into the standard form \eqref{eq:KdV}.

\end{example}

\subsection{Equivalence Transformations and Point Symmetries}\label{sec:eqtr:ps}
Lie groups of equivalence transformations \eqref{eq:ch1:symm:equivalence:eq_trans} are closely related to Lie groups of point symmetries of differential equations (e.g., \cite{BCABook, ovsiannikov2014group}). Let  $\PDEs{R}{x}{u} \in \mathcal{F}_K$ given by \eqref{eq:ch1:sym:PDEsys_eqivtr} be a specific DE system within the indicated family, for a fixed set of constitutive quantities $K$.

Consider a one-parameter Lie group of point transformations
\beq\label{eq:ch1:sec11:PDE_symms:point}
\begin{array}{lll}
(x^*)^i&=&f^i(x,u;\eps)=x^i + \eps \xi^i(x,u) + O(\eps^2),\quad i=1,\ldots ,n,\\[2ex]
(u^*)^\mu&=&g^\mu(x,u;\eps)=u^\mu + \eps \eta^\mu(x,u) + O(\eps^2),\quad \mu=1,\ldots ,m,
\end{array}
\eeq
with the infinitesimal generator
\beq\label{eq:ch1:sec11:LiePointSymAlg:X}
\sg{X} = \xi^i(x,u)\frac{\partial }{\partial x^i}+\eta^\mu(x,u)\frac{\partial }{\partial u^\mu},
\eeq
and the $k$-th prolongation of \eqref{eq:ch1:sec11:LiePointSymAlg:X} is given by
\beq\label{eq:ch1:sec11:LiePointSymAlg:Xk}
\barr
\sg{X}^{(k)} = &\xi^i(x,u)\dfrac{\partial }{\partial x^i}+\eta^\mu(x,u)\dfrac{\partial }{\partial u^\mu}+
\eta^{(1)\,\mu}_{i}(x,u,\partial u) \dfrac{\partial }{\partial u^\mu_{i}}\\[\dfracskip]
&+\ldots + \eta^{(k)\,\mu}_{i_1\ldots i_k}(x,u,\partial u,\ldots,\partial^k u)\dfrac{\partial }{\partial
u^\mu_{{i_1}\ldots {i_k}}}.
\earr
\eeq
Here the extended infinitesimals $\eta^{(1)\,\mu}_{i}$, $\ldots $, $\eta^{(k)\,\mu}_{i_1\ldots i_k}$ are defined through the formulas
\beq\label{eq:ch1:sec11:higher_etas1N}
\eta^{(1)\,\mu}_{i}= \sg{D}_{i}\eta^\mu - (\sg{D}_{i}\xi^j)u^\mu_{j},\qquad \eta^{(k)\,\mu}_{i_1\ldots i_k}= \sg{D}_{i_k}\eta^{(k-1)\,\mu}_{i_1\ldots i_{k-1}} -
(\sg{D}_{i_k}\xi^j)u^\mu_{i_1\ldots i_{k-1} j},
\eeq
for $\mu=1,\ldots ,m$, and $i, i_p=1,\ldots ,n$ for $p=1,\ldots ,k$ (e.g., \cite{hydon2000symmetry, Olver, BCABook}).

\begin{definition} \label{def:symm:pt}
A one-parameter Lie group of point transformations \eqref{eq:ch1:sec11:PDE_symms:point} acting on the independent and dependent variables of the DE system $\PDEs{R}{x}{u}$ \eqref{eq:ch1:sym:PDEsys_eqivtr} is a \emph{one-parameter Lie group of point symmetries} of this system if and only if its $k$th extension \eqref{eq:ch1:sec11:LiePointSymAlg:Xk} leaves invariant the solution manifold of
$\PDEs{R}{x}{u}$  in $(x,u,\partial u, \ldots,\partial^k u)$-space.
\end{definition}

The notion of a point symmetry naturally extends to include higher-order local symmetries, nonlocal symmetries, approximate symmetries, and other related objects (see, e.g., \cite{BlumanKumei, BCABook, ibragimov1995crc1, ibragimov1995crc3, cheviakov2007gem, cheviakov2010symbolic}). The symmetry structure of a DE system contains essential information about the properties of that system, and thus can vary greatly. For example, differential equations exist that have no point symmetries, a several-parameter group of point symmetries, or an infinite number of point symmetries. Infinite-dimensional  groups of point symmetries arise when the symmetry components $\xi^i, \eta^\mu$ in \eqref{eq:ch1:sec11:LiePointSymAlg:X} involve arbitrary functions. Such symmetries are particularly useful for the construction of families of new exact solutions from known ones, as well as other applications. Moreover, for a nonlinear DE system with an infinite point (or nonlocal) symmetry group, one can determine whether or not this system can be mapped into a linear system by an invertible (or, respectively, a non-invertible) transformation, and find an explicit form of such a transformation (\cite{BlumanKumei, BCABook} and references therein).

\begin{remark} For families of DE systems, \emph{classifications} of their analytical properties with respect to constitutive elements are of high theoretical and practical importance. Examples of such problems include local and nonlocal symmetry and conservation law classifications. Classification tables are usually presented modulo known equivalence transformations, i.e., only for forms of constitutive functions and/or parameters that are not related by an equivalence transformation. Such classifications are available for multiple families of equations of mathematical physics; see, for instance, \cite{ibragimov1995crc1,BCABook}.

Equivalence transformations can be used directly to simplify symmetry determining equations during symbolic computations. A symbolic software package \verb|SymmetryClassification| for \verb|Maple| developed by Huang and Lisle includes a \verb|detEqsForEquiv| routine which computes rather a restricted class of point equivalence transformations, namely, ones where components for each variable or arbitrary element depend only on that specific variable or arbitrary element. A reduced simplified version of the \verb|rifsimp|-generated symmetry classification tree is consequently obtained \cite{Huang2009,lisle2010algorithmic}. The work contains detailed examples of symmetry classification for a family of nonlinear heat equations, and a family of nonlinear convection-diffusion (Richards) equations.
\end{remark}

\begin{remark}
The following obvious relationships hold between the group of equivalence transformations of a DE family \eqref{eq:ch1:sym:PDEsys_eqivtr} and the groups of point symmetries of its members:
\begin{itemize}
  \item A point symmetry \eqref{eq:ch1:sec11:PDE_symms:point} of a DE system $\PDEs{R}{x}{u} \in \mathcal{F}_K$ is an equivalence transformation of the family
$\mathcal{F}_K$ \eqref{eq:ch1:sym:PDEsys_eqivtr} if \eqref{eq:ch1:sec11:PDE_symms:point} is point symmetry for all systems in $\mathcal{F}_K$.
  \item An equivalence transformation \eqref{eq:ch1:symm:equivalence:eq_trans} of the DE family $\mathcal{F}_K$ \eqref{eq:ch1:sym:PDEsys_eqivtr} is a point symmetry of its every member if and only if it does not involve components corresponding to the arbitrary elements of the family.
\end{itemize}
\end{remark}

In practice, one-parameter Lie group of point symmetries are found from the following \emph{invariance condition} (e.g., \cite{ovsiannikov2014group, Olver, hydon2000symmetry, BCABook}).
\begin{theorem} 
Consider a one-parameter Lie group of point transformations \eqref{eq:ch1:sec11:PDE_symms:point}, with the corresponding infinitesimal generator \eqref{eq:ch1:sec11:LiePointSymAlg:X} and its
$k$th extension \eqref{eq:ch1:sec11:LiePointSymAlg:Xk}. Then the transformation \eqref{eq:ch1:sec11:PDE_symms:point} is a point symmetry of a nondegenerate DE
system $\PDEs{R}{x}{u}$ \eqref{eq:ch1:sym:PDEsys_eqivtr} if and only if for each $\alpha=1,\ldots ,N$,
\beq\label{eq:ch1:sec11:PDE_symms:Pointsym:deteq1}
\sg{X}^{(k)} R^\alpha(x,u,\partial u, \ldots , \partial^k u)=0,
\eeq
whenever
\beq\label{eq:ch1:sec11:PDE_symms:Pointsym:deteq2}
R^\sigma(x,u,\partial u, \ldots , \partial^k u)=0, \quad \sigma=1,\ldots ,N.
\eeq
\end{theorem}

The computation of point symmetries from the determining equations \eqref{eq:ch1:sec11:PDE_symms:Pointsym:deteq1}, \eqref{eq:ch1:sec11:PDE_symms:Pointsym:deteq2} proceeds through the Lie's algorithm:
\begin{enumerate}
  \item Obtain the \emph{determining equations} by evaluating \eqref{eq:ch1:sec11:PDE_symms:Pointsym:deteq1} on the solution manifold of the given PDE system, i.e., through the
  substitution of the DEs \eqref{eq:ch1:sec11:PDE_symms:Pointsym:deteq2} and  their differential consequences into the invariance condition \eqref{eq:ch1:sec11:PDE_symms:Pointsym:deteq1}.
  \item Obtain the \emph{split system of determining equations}, using the fact that the symmetry components $\xi^i, \eta^\mu$ do
      not depend on the derivatives of $u$: in the determining equations, set to zero all coefficients at all independent combinations of
      derivatives of dependent variables.
\end{enumerate}
The application of the Lie's algorithm and its extensions to nontrivial DE systems often requires
extensive algebraic manipulation. Indeed, split symmetry determining equations are linear overdetermined PDE systems for the unknown symmetry components $\xi^i, \eta^\mu$; in practice, such systems may involve from a dozen to hundreds or thousands of equations. To treat such huge systems, symbolic computer software is commonly used (see, e.g., \cite{cheviakov2007gem, cheviakov2010symbolic}). In particular, methods based on differential Gr\"obner bases and the characteristic set method are successfully used in computer algebra packages \cite{lisle1992equivalence, reid1996reduction, wolf2002crack, chaolu2010algorithm}. A review of relevant work and symbolic software can be found in \cite{BCABook} (Section  5.1).


In practical computations, groups of equivalence transformations \eqref{eq:ch1:symm:equivalence:eq_trans} of DE families are systematically sought using the above Lie's algorithm, similarly to finding Lie groups of point symmetries, with arbitrary elements treated as dependent variables, and the dependence of the transformation components restricted accordingly. Details are given in Section \ref{sec:algo:equi} below.

\subsection{Extensions of the Notion of Equivalence Transformations}\label{subsec:eq:tr:gen}

Definition \ref{def:equivalence:std} can be extended in several different ways. Details and further examples of such extensions can be found in \cite{lisle1992equivalence, meleshko1996generalization, ivanova2005conservationRD, vaneeva2009enhanced,  popovych2010admissible} and references therein, and are out of scope of this paper. We briefly review several generalizations of equivalence transformations that are relevant to systematic symbolic computations.

A general approach to the extension of the notion of one-parameter Lie groups of point equivalence transformations \eqref{eq:ch1:symm:equivalence:eq_trans} can be formulated in terms of seeking transformations of the form
\beq \label{eq:eqtr:mostgen}
\begin{array}{llll}
(x^*)^i&=&f^i[x,u,K],\quad i=1,\ldots ,n,\\
(u^*)^\mu&=&g^\mu[x,u,K],\quad \mu=1,\ldots ,m,\\
(K^*)^\ell&=&G^\ell[x,u,K], \quad \ell=1,\ldots, L,
\end{array}
\eeq
that map PDE systems in the family $\mathcal{F}_K$ into PDE systems in the same family. Generally, transformations \eqref{eq:eqtr:mostgen} may be neither point, nor local transformations; the bracket notation in \eqref{eq:eqtr:mostgen} denotes possible local or nonlocal dependencies of the transformation components on the independent and dependent variables and arbitrary elements, as well as, possibly, on derivatives of the dependent variables.

The following specific instance of general transformations \eqref{eq:eqtr:mostgen} have been considered in the literature.






\bigskip\noindent\textbf{(A) Generalized equivalence transformations.} An extension of one-parameter Lie groups of point equivalence transformations \eqref{eq:ch1:symm:equivalence:eq_trans}, in particular, Lie groups with the corresponding infinitesimal generators
\beq\label{eq:ext:Melesh}
\sg{X} = \xi^i(x,u,K)\frac{\partial}{\partial x^i}+\eta^\mu(x,u,K)\frac{\partial }{\partial u^\mu} + \zeta^\nu(x,u,K)\frac{\partial }{\partial K^\nu},
\eeq
were defined in \cite{meleshko1996generalization} (cf. \cite{lisle1992equivalence}), and are referred to as groups of ``generalized equivalence transformations". The essential difference here is that the transformation components for the independent and/or dependent variables $\xi^i(x,u,K)$, $\eta^\mu(x,u,K)$ may additionally depend on arbitrary elements $K$. Examples of such transformations are known in the literature. For example, they were computed in \cite{popovych2004group} for a class of nonlinear (1+1)-dimensional Schr{\"o}dinger equations with a potential $V(x,t)$ and a power nonlinearity $\gamma=\const$ given by
\[
i\psi_t +\psi_{xx}+ |\psi|^\gamma\psi + V(x,t)\psi=0,
\]
For an infinite-parameter group of generalized equivalence transformations derived in that work, transformations for the dependent variable $\psi(x,t)$ depend on the arbitrary constant parameter $\gamma$.

\begin{remark} \label{rem:mel:false}
It is also worth mentioning that Lie groups of generalized equivalence transformations \eqref{eq:ext:Melesh} where the components $\xi^i(x,u,K)$ and/or $\eta^\mu(x,u,K)$ of the independent and/or dependent variables would depend on \emph{arbitrary functions} present in the given system, rater than \emph{arbitrary constant parameters}, reliable examples of such verified examples have not appeared in the literature been available to this day. Despite multiple re-citations, the example given in the original work \cite{meleshko1996generalization} does not seem to be correct, as can be verified by a direct substitution. The first and only example of the indicated kind the author is aware of is due to Opanasenko, Bihlo and Popovych, appearing in their yet unpublished work \cite{Opanasenko}. Consider the KdV-Burgers PDE family for the unknown function $u(x,t)$ given by
\beq\label{eq:KdVB}
u_t + C u u_x = \sum_{k=0}^r A_k u_k + B,
\eeq
where $u_k\equiv \partial^k u /\partial x^k$, $k=1,\ldots,r$, $r\geq 2$ is a fixed integer, and $A_k=A_k(x,t)$, $B=B(x,t)$, $C=C(x,t)$ are constitutive functions (arbitrary elements). The work \cite{Opanasenko} contains several statements concerning the structure of the generalized equivalence transformation \emph{groupoid} of the PDE family \eqref{eq:KdVB}. In particular, equivalence transformations may be used to invertibly map any PDE within \eqref{eq:KdVB} into one with $A_r=1$ and $A_1=0$. Moreover,  in \cite{Opanasenko},
it is shown that when $A_r=1$ and $A_1=0$, the family \eqref{eq:KdVB} admits the equivalence transformations with the transformed variables given by
\beq\label{eq:KdVB:eqivtr}
x_* = F_1x+F_0, \qquad t_* = G,\qquad u_*(x_*,t_*) = U_1u+U_0,\\[3ex]
\eeq
where $F_0=F_0(t)$, $G=G(t)$ and $U_1=U_1(t)$ are arbitrary functions, $F_1=F_1(t)$ is related to $G(t)$, $U_0$ is given by
\beq\label{eq:KdVB:eqivtr2}
U_0=U_0(x,t)=\dfrac{F_1' x +F_0'}{F_1 C}U_1,
\eeq
and the transformations for $A_k, B, C$ are given by complex formulas found in \cite{Opanasenko}.

Importantly, the transformation of the dependent variable $u(x,t)$  \eqref{eq:KdVB:eqivtr2} involves the arbitrary element $C(x,t)$. These transformations are interpreted as the generalized equivalence group
of the PDE family \eqref{eq:KdVB} if one formally extends the set of arbitrary elements to include the time derivative $C_t$ as well as $x$-derivatives of $C$ of orders $1,\ldots, r$. The transformations \eqref{eq:KdVB:eqivtr} do not arise as point generalized equivalence transformations of the PDE family \eqref{eq:KdVB} without such prolongation.

The above example is readily verified using \verb|GeM| symbolic software discussed in this contribution (see Sections \ref{sec:algo:gem} and \ref{sec:eg:BKdV} below).

%
%
%
%

\end{remark}



\begin{remark}
One may also note that it is possible for the equivalence transformations to involve constant arbitrary elements, and therefore belong to the class \eqref{eq:eqtr:mostgen}, even when these transformations are point equivalence transformations \eqref{eq:ch1:symm:equivalence:eq_trans}. An instance of such a transformation is in fact contained in Example \ref{eg:2:2} above. Indeed, in the equivalence transformations \eqref{eq:KdV:eqivtr}, the choice $A_3=a$ in \eqref{eq:dimKdV:trans:KillUx:2} corresponds to a generalized equivalence transformation, since $a$ is a constant constitutive parameter.
\end{remark}

\bigskip\noindent\textbf{(B) Generalized extended equivalence transformations.}
One can consider further extensions of the notion of equivalence transformations, for example, the ones involving nonlocal transformations of variables and/or constitutive components. For example, \emph{generalized extended equivalence groups} include transformations of the form \eqref{eq:eqtr:mostgen},
where the expressions for the transformed quantities are not necessarily point, but become a point transformation with respect to the independent and dependent variables after fixing the values of the arbitrary elements. Examples of transformations \eqref{eq:eqtr:mostgen} involving arbitrary functions that arise as solutions of ordinary differential equations are contained, for instance, in \cite{vaneeva2009enhanced}; we present an example from that paper below.

\begin{example} \label{eg:geneqtr:Vaneeva} Consider the PDE family
\beq\label{eq:Vaneeva:sys}
f(x)u_t = (g(x)u_x )_x +h(x)u^m,
\eeq
where $f(x)$, $g(x)$, $h(x)$ are arbitrary smooth constitutive functions satisfying $f(x)g(x)h(x) \ne 0$, and  $m$ is a constant constitutive parameter. It is shown in \cite{vaneeva2009enhanced}, for example, that the transformations
\beq\label{eq:Van:et1}
\barr
x^* = \phi(x),\quad t^* = \delta_1 \,t + \delta_2,\quad u^* = \psi(x) \,u,\quad m^* = m,\\[2ex]
f^* = \dfrac{\delta_0\,\delta_1}{\phi_x\,\psi^2}\,f,\quad g^* = \dfrac{\delta_0\,\phi_x}{\psi^2}\,g,\quad h^* = \dfrac{\delta_0}{\phi_x\,\psi^{m+1}}\,h
\earr
\eeq
map the PDE family \eqref{eq:Vaneeva:sys} into itself, for any choice of smooth $\phi(x)\ne0$, constants $\delta_0$, $\delta_1$, $\delta_2$ satisfying $\delta_0\,\delta_1\ne0$, if $\psi(x)$ is an arbitrary solution of an ODE
\[
\left(\dfrac{g\,\psi_x}{\psi^2}\right)_x=0.
\]
\end{example}

Generalized extended equivalence transformations similar to \eqref{eq:Van:et1} have been extensively used in symmetry and conservation law classification problems in \cite{ivanova2005conservationRD, vaneeva2009enhanced, ivanova2010groupI, vaneeva2012extended} and other works of the same group. In fact, the following early example of generalized extended equivalence transformations appears in \cite{varley1985exact}, and is discussed in detail in \cite{lisle1992equivalence}.

\begin{example} \label{eg:geneqtr:Lisle} Consider a family of PDE systems of nonlinear telegraph equations given by
\beq \label{eq:NLT:eg:Lisle:equiv:System}
v_{t}=u_x,\qquad v_{x}=c^2(u)\,u_t+b(u),
\eeq
with dependent variables $u=u(x,t)$, $v=v(x,t)$ and two arbitrary functions $b(u)$, $c(u)$. Then for any fixed value of the parameter $\eps$, the change of variables
\beq \label{eq:NLT:eg:Lisle:equiv:Tr}
\barr
x^*=x-\eps v,& v_{t}=u_x,\qquad t^*=t - \eps \displaystyle \int_{u_0}^u \dfrac{c^2(z)}{1- \eps b(z)}\,dz\\[2ex]
u^*=u, & v^*=v
\earr
\eeq
maps each nonlinear telegraph system to another system from the same family, with new arbitrary functions given by
\beq \label{eq:NLT:eg:Lisle:equiv:Tr:BC}
b^*(u^*)=\dfrac{b(u)}{1- \eps b(u)},\qquad c^*(u^*)=\dfrac{c(u)}{1- \eps b(u)}.
\eeq
The transformations defined by \eqref{eq:NLT:eg:Lisle:equiv:Tr}, \eqref{eq:NLT:eg:Lisle:equiv:Tr:BC} are nonlocal transformations that do not form a Lie group of point transformations in the $(x, t, u, v)$ space. Indeed, the transformed quantities $x^*, t^*, u^*, v^*$ depend not only on
the values of the independent and dependent variables $x, t, u, v$, but also
on an \emph{a priori} specification of the arbitrary elements $b(u)$, $c(u)$, and hence belong to the class of generalized extended equivalence transformations.
\end{example}

\begin{remark}
From the computational point of view, it is important to remark that in general, the generalized extended equivalence transformations do not form a Lie group of point transformations even in the space of variables extended by arbitrary elements. For example, the transformations defined by \eqref{eq:NLT:eg:Lisle:equiv:Tr}, \eqref{eq:NLT:eg:Lisle:equiv:Tr:BC} do not form a Lie group in the space of $(x, t, u, v, b, c)$, and hence do not belong to the set of generalized equivalence transformations \eqref{eq:ext:Melesh}.
\end{remark}

\bigskip\noindent\textbf{(C) Potential equivalence transformations.} Another example of transformations which are outside of the class of local Lie groups of equivalence transformations \eqref{eq:ch1:symm:equivalence:eq_trans} is given by potential equivalence transformations. Such nonlocal transformations can arise as Lie groups of point equivalence transformations of a \emph{potential system}. Consider a family of nonlinear wave equations in (1+1) dimensions \cite{bluman2007nonlocally, BCABook} given by
\beq
\label{eq:ch4:nonl_wave_ch4} u_{tt}=(c^2(u)u_x)_x,
\eeq
for the unknown $u=u(x,t)$, with an arbitrary constitutive function $c(u)$. One can show that the  infinitesimal generators of the group of point equivalence transformations of the family \eqref{eq:ch4:nonl_wave_ch4} are given by
\beq\label{eq:wav:ET:U}
\barr
\sg{Z}_1=\dfrac{\partial }{\partial t},\quad \sg{Z}_2=\dfrac{\partial }{\partial x},\quad \sg{Z}_3=\dfrac{\partial }{\partial u},\quad \sg{Z}_4=t \dfrac{\partial }{\partial u},\\[3ex]
\sg{Z}_5=u\dfrac{\partial }{\partial u}, \quad
\sg{Z}_6=t\dfrac{\partial }{\partial t} + x\dfrac{\partial }{\partial x},\quad
\sg{Z}_7=t\dfrac{\partial }{\partial t} - c \dfrac{\partial }{\partial c},
\earr
\eeq
and the global group has the form
\begin{gather}
{x}^*=a_6{x}+a_2,\quad {t}^*=a_6a_7{t}+a_1,\quad {u}^*=a_5 u +a_4 t + a_3,\quad
{c}^*({u}^*)=a_7^{-1}{c}({u}),\label{eq:ch4:nonl_wave_equivt}
\end{gather}
where $a_1, \ldots, a_7$ are arbitrary constants with $a_5 \,a_6 \,a_7\ne0$. Additional  equivalence transformations arise as local equivalence transformations of a PDE system nonlocally related to \eqref{eq:ch4:nonl_wave_ch4} \cite{BC1,BC2,bluman2007nonlocally, BCABook}. In particular, for an arbitrary $c(u)$, PDEs of the family \eqref{eq:ch4:nonl_wave_ch4} admit local conservation laws corresponding to multipliers $1,t,x,xt$. For the multiplier $t$, the conservation law is
\beq\label{eq:UW:CL}
\dfrac{\partial}{\partial t}(t u_t-u) - \dfrac{\partial}{\partial x} (t c^2(u) u_x) =0.
\eeq
The corresponding family of potential systems with a nonlocal variable $w(x,t)$, is given by
\beq \label{eq:ch4:nonl_wave_ch4:w}
w_{x}=tu_t-u,\qquad w_{t}=tc^2(u)u_x;
\eeq
every PDE of this family is nonlocally related and equivalent (in a non-invertible manner) to a PDE  of the family \eqref{eq:ch4:nonl_wave_ch4}. A computation of Lie groups of point equivalence transformations of the family of potential systems \eqref{eq:ch4:nonl_wave_ch4} yields the eight-dimensional Lie algebra of generators
\beq\label{eq:wav:ET:UW}
\barr
\sg{W}_i=\sg{Z}_i, \quad i=2, 4, 7,\\[1.5ex]
\sg{W}_1=\dfrac{\partial }{\partial w},\qquad \sg{W}_3=\sg{Z}_3-x\dfrac{\partial }{\partial w},\\[3ex]
\sg{W}_5=\sg{Z}_5-t\dfrac{\partial }{\partial t}-x\dfrac{\partial }{\partial x}, \qquad
\sg{W}_6=\sg{Z}_6+w\dfrac{\partial }{\partial w},\\[3ex]
\sg{W}_8=tu\dfrac{\partial }{\partial t}+w\dfrac{\partial }{\partial x} + u^2 \dfrac{\partial }{\partial u} -2uc\dfrac{\partial }{\partial x}.
\earr
\eeq
In particular, the time translation $\sg{Z}_1$ disappears, and two new generators, the obvious $\sg{W}_1$ and an additional generator $\sg{W}_8$, appear as local equivalence transformation generators of the potential system \eqref{eq:ch4:nonl_wave_ch4:w}.

The global group has a rather complicated form; it is simpler to write it in two parts. The one corresponding to the first seven generators is given by
\begin{equation} \label{eq:ch4:NLW:UW_equivt1}
\begin{array}{rll}
{x}^*&=&A_5^{-1} A_6x+A_2,\quad {t}^*=A_5^{-1} A_6A_7t,\quad {u}^*=A_5u+A_4t+A_3,\\[1ex]
{w}^*&=&A_6w-A_3A_5^{-1} A_6x+A_1,\quad {c}^*({u}^*)=A_7^{-1}c(u)
\end{array}
\end{equation}
(cf. \cite{bluman2007nonlocally, BCABook}), where $A_1, \ldots, A_7$ are arbitrary constants with $A_1 A_2 A_3\ne0$, corresponding to the generators $\sg{W}_1, \ldots, \sg{W}_7$. The global group corresponding to the infinitesimal generator $\sg{W}_8$ yields a one-parameter group of transformations
\begin{equation} \label{eq:ch4:NLW:UW_equivt2}
\barr
{x}^* = x-B w,\qquad {t}^*=\dfrac{t}{1+B{u}},\qquad {u}^*=\dfrac{u}{1+B{u}},\\[2.5ex]
{w}^*=w, \qquad {c}^*({u}^*) = (1+Bu)^2\,c(u)
\earr
\end{equation}
with a parameter $B$.
The equivalence transformations \eqref{eq:ch4:NLW:UW_equivt2} are nonlocal projective-type transformations of the nonlinear wave equation family \eqref{eq:ch4:nonl_wave_ch4}; they have been employed in a symmetry classification problem \cite{bluman2007nonlocally}. Further examples of nonlocal equivalence transformation can be found in, e.g.,  \cite{BCABook,lisle1992equivalence}.

In Section \ref{sec:eqtr:NLWave} below, the symbolic software package \verb|GeM| is used to compute the eight generators \eqref{eq:wav:ET:UW} and the corresponding global groups \eqref{eq:ch4:NLW:UW_equivt1}, \eqref{eq:ch4:NLW:UW_equivt2}.

\subsection{Systematic Computation of Lie Groups of Generalized Equivalence Transformations}\label{sec:algo:equi}

For a given family $\mathcal{F}_K$ of DE systems $\PDEs{R}{x}{u; K}$ \eqref{eq:ch1:sym:PDEsys_eqivtr},
the following algorithm can be used for the computation of (generalized) point equivalence transformations \eqref{eq:ch1:symm:equivalence:eq_trans}, \eqref{eq:ext:Melesh}.

If the set $K$ involves arbitrary functions, it is assumed that these are functions of only the components of $x,u$, and are not functions of, for example, derivatives of $u$.

\begin{enumerate}
  \item Replace the constitutive functions and/or parameters $K=(K^1,\ldots,\,K^L)$ by new dependent variables $(K^1(x),\ldots,\,K^L(x))$.
  Thus consider a new DE system $\PDEs{\widetilde{R}}{x}{u, K}$ with $m+L$ dependent variables and no arbitrary elements.

  \item Seek point symmetries of $\PDEs{\widetilde{R}}{x}{u, K}$, with infinitesimal generators
\beq\label{eq:EQ:X}
\sg{X} = \xi^i(x,u,K)\frac{\partial }{\partial x^i}+\eta^\mu(x,u,K)\frac{\partial }{\partial u^\mu}+\theta^\lambda(x,u,K)\frac{\partial }{\partial K^\lambda}.
\eeq

  \item Obtain the \emph{split system of determining equations} for $\PDEs{\widetilde{R}}{x}{u, K}$, as outlined in Section \ref{sec:eqtr:ps}.

  \item If the arbitrary elements $K$ of the original family $\PDEs{R}{x}{u; K}$ \eqref{eq:ch1:sym:PDEsys_eqivtr} of DE systems contained arbitrary functions, introduce restrictions of the form
\beq\label{eq:EQ:restric}
\dfrac{\partial \xi^i(x,u,K)}{\partial K^\gamma}=0, \quad \dfrac{\partial \eta^\mu(x,u,K)}{\partial K^\delta}=0,\quad \dfrac{\partial \theta^\lambda(x,u,K)}{\partial x^j}=0,
\eeq
etc., as appropriate, to exclude the dependence of transformation components of the arbitrary elements on variables they do not depend on, as well as the dependence of transformation components of the variables of the system on inappropriate arbitrary elements. For example, for the DE family \eqref{eq:1}, the infinitesimal generator of the generalized equivalence transformations has the form
\beq\label{eq:EQ:X:wave}
\sg{X} = \xi(x,t,u,c)\frac{\partial }{\partial x}+\tau(x,t,u,c)\frac{\partial }{\partial t}+\eta(x,t,u,c)\frac{\partial }{\partial u}+\theta(x,t,u,c)\frac{\partial }{\partial c},
\eeq
and the transformation for $c(u)$ must not explicitly depend on the variables $x,t$. Therefore the restrictions on the component $\theta$ are given by
\beq\label{eq:EQ:restric:Wave}
\dfrac{\partial \theta(x,t,u,c)}{\partial x}=\dfrac{\partial \theta(x,t,u,c)}{\partial t}=0.
\eeq

\item In order to simplify computations, additional restrictions can be introduced at this stage, for example,
\beq\label{eq:EQ:restric:voluntary}
\dfrac{\partial \xi^i(x,u,K)}{\partial K^\gamma}=0,\quad i=1,\ldots, n,\quad \gamma=1,\ldots, L,
\eeq
if the transformations for the independent variables are assumed to be independent of the arbitrary elements.


\item Append all restrictions, in the form of linear PDEs, to the split system of determining equations.

  \item Simplify and solve the augmented split system of determining equations, to find the infinitesimal generators \eqref{eq:EQ:X} of the equivalence transformations.

  \item Integrate to obtain the global group.  For each infinitesimal generator \eqref{eq:EQ:X}, the corresponding one-parameter Lie group of equivalence transformations \eqref{eq:ch1:symm:equivalence:eq_trans} is found through the solution of the initial-value problem
\beq\label{eq:EQ:Lie1}
\barr
\dfrac{d }{d\eps} (x^*)^i=\xi^i(x^*,u^*, K^*),\quad i=1,\ldots, n,\\[2ex]
\dfrac{d }{d\eps} (u^*)^\mu=\eta^\mu(x^*,u^*, K^*),\quad i=1,\ldots, m,\\[2ex]
\dfrac{d }{d\eps} (K^*)^\lambda=\theta^\lambda(x^*,u^*, K^*),\quad i=1,\ldots, L,\\[2ex]
(x^*)^i|_{\eps=0}=x^i,\quad (u^*)^\mu|_{\eps=0}=u^\mu,\quad  (K^*)^\lambda|_{\eps=0}= K^\lambda,
\earr
\eeq
where $\eps$ is the group parameter.

\end{enumerate}

\begin{remark}
The above procedure is readily generalized to the case when the given DE system involves both an arbitrary function and its derivative(s) (for example, $c(u)$ and $c'(u)$). In such cases, in Step 1 of the above algorithm, both the arbitrary function at hand and its derivative(s) can be replaced by different additional dependent variables (for example, $C(x,t)$ and $D(x,t)$).
\end{remark}


\section{Symbolic Computation of Equivalence Transformations}\label{sec:algo:gem}

The symbolic software package \verb|GeM| for \verb|Maple| contains routines for point and local (higher-order) symmetry analysis of systems of ordinary and partial differential equations, routines for local conservation law analysis, and other related functions. For details on the available routines, additional options, methods, background, and examples, see \cite{cheviakov2007gem, cheviakov2010symbolic, GemReferenceOnline}.

The \verb"GeM" package uses an efficient representation of the DEs  and resulting symmetry/conservation law
determining equations, through the treatment of dependent variables and their derivatives as \verb"Maple" symbols, rather
than functions or expressions: $\partial B_1 /\partial x\equiv \verb"B1x"$, etc. For example, the
\verb"Maple" representation for the spatial divergence of a 3-vector $(B_1(x,y,z)$, $B_2(x,y,z)$,$B_3(x,y,z))$ in \verb"GeM" routines is
\begin{equation}\label{eq:B123}
{\frac {\partial }{\partial x}}{B_1}(x,y,z) +{\frac {
\partial }{\partial y}}{B_2}(x,y,z) +{\frac {
\partial }{\partial z}}{B_3}(x,y,z)= \verb"B1x + B2y + B3z".
\end{equation}
In the implementation of recent versions of the \verb|GeM| package, its author has made a conscious choice to move away from the standard \verb|Maple| package structure, keeping \verb|GeM| a collection of routines and variable definitions. This structure provides direct access to all variables and internal routines for an advanced user, while the chosen naming convention prevents the clash with  user variables and functions.

The \verb|GeM| package can be used to automate computations of Lie groups of point and generalized equivalence transformations. The sequence of steps for such computations, following the algorithm of Section \ref{sec:algo:equi}, is outlined below. We note that the algorithm described in the current work can be applied to a given family of DE systems \eqref{eq:ch1:sym:PDEsys_eqivtr} as it stands, or to a related system (such as a potential system or a nonlocally related subsystem \cite{BC1}), in order to seek a more general (for example, nonlocal) equivalence transformations.

In the subsequent Section, examples of symbolic computations are presented.

\bigskip\noindent \textbf{Step A. Initialization.} Clear the variables. Initialize the package. If required, also initialize a sub-package for global group computations.
\beq\label{eq:gem:initeq}
\barr
\verb|restart:|\\
\verb|read("d:/gem32_12.mpl"):|\\
\verb|read("d:/gem_globgroup_and_equiv.txt"):|
\earr
\eeq

\bigskip\noindent \textbf{Step B. Declare variables.} The user creates a collection of independent variables called, for example, \verb|ind|, and a collection of dependent variables and constitutive elements (which for the purpose of equivalence transformation computations are treated as dependent variables) called, for example,  \verb|dep|, and declares them.
\[
\barr
\verb|gem_decl_vars(indeps=[ind], deps=[dep]);|
\earr
\]
The readability of names of variables and their derivatives is enhanced when one employs small Latin letters to denote the independent variables, and capital Latin letters for the dependent variables (cf. \eqref{eq:B123}).

\bigskip\noindent \textbf{Step C. Declare the DEs.} The user defines differential equations named, for example, \verb|Eq1| and \verb|Eq2|, and declares them.
\[
\barr
\verb|gem_decl_eqs([Eq1,Eq2], solve_for=[LeadDer]);|
\earr
\]
The equations need to be solvable algebraically with respect to the specified leading derivatives \verb"LeadDer" specified in the list \verb"solve_for". Within the same routine, necessary differential consequences of the DEs are automatically computed and solved with respect to the differential consequences of the leading derivatives.

\bigskip\noindent \textbf{Step D. Generate the determining equations.}  Next, one generates the equivalence transformation determining equations (Section \ref{sec:algo:equi}, Step 3), using the command
\[
\verb"det_eqs:=gem_symm_det_eqs([ind,dep]);"
\]
where the list \verb|[ind,dep]| denotes the common dependence of all components of the generator \eqref{eq:EQ:X}.

The routine \verb|gem_symm_det_eqs| outputs a set containing the overdetermined split linear system of determining equations for the equivalence transformation tangent vector field components $\xi^i(x,u,K)$, $\eta^\mu(x,u,K)$, $\theta^\lambda(x,u,K)$. In the \verb|GeM| notation, these components are denoted by symbols like \verb|xi_x| (for an independent variable $x$) and \verb|eta_U| (for a independent variable $u$).

In order to list the tangent vector field components, one can use the routine
\[
\barr
\verb|OFF:|\\
\verb|sym_components:=gem_symm_components();|\\
\verb|ON:|
\earr
\]
Here the \verb|OFF:| and \verb|ON:| commands are used to switch off and on the output suppression of the function dependencies in  \verb|Maple|. It is done here to force \verb|Maple| to print the equivalence transformation components explicitly, with their dependencies.

\bigskip\noindent \textbf{Step E. Restrict dependencies.} As per Section \ref{sec:algo:equi}, Step 4, one can restrict the sets of variables the equivalence transformation tangent vector field components depend on, by appending linear differential equations like \eqref{eq:EQ:restric} to the determining equations. Restriction equations are generated using the command
\[
\barr
\verb|restriction_eqs:=gem_generate_EquivTr_dependence([|\\
\verb|     [[<variables1>],[<dep1>]],|\\
\verb|     [[<variables2>],[<dep2>]],|\\
\verb|     ...,|\\
\verb|]);|
\earr
\]
This will cause the equivalence transformation  component(s) listed in a comma-separated sequence \verb|<variables1>| to depend on variables listed in a comma-separated sequence \verb|<dep1>|, etc. The output stored in the  variable \verb|restriction_eqs| contains a \verb|Maple| set of linear differential restrictions in the form \eqref{eq:EQ:restric}.

The augmented split set of equivalence transformation determining equations is then constructed using a set union
\[
\verb|det_eqs_all:=det_eqs union restriction_eqs:|
\]

\bigskip\noindent \textbf{Step F. Rif-simplify the augmented determining equations.} At this optional step, one uses the \verb|rifsimp| routine \cite{reid1996reduction} for the Gr\"obner basis-based reduction of the overdetermined linear system of determining equations obtained at the previous step.
\[
\barr
\verb|simplified_eqs := DEtools[rifsimp](det_eqs_all, sym_components, |\\
\verb|     casesplit, mindim = 1);|
\earr
\]
The current step is particularly important for large systems of determining equations, since through the elimination of the redundancy, it commonly leads to a drastic simplification and reduction of the number of PDEs in the system. The output of \verb|rifsimp| is a \verb|Maple| table, with the simplified set of equations found in \verb|simplified_eqs[Solved]|.

The \verb|rifsimp| routine is capable of the solution space dimension prediction (when used with \verb|mindim = 1| option), and case splitting/classification (when used with \verb|casesplit| option); for details, see \cite{cheviakov2007gem, cheviakov2010symbolic, GemReferenceOnline} or \verb|Maple| help.

\bigskip\noindent \textbf{Step G. Solve the determining equations.} The solution as a set is obtained, for example, by  \verb|Maple|  \verb|pdsolve| routine:
\[
\verb|eq_tr_sol:=pdsolve( simplified_eqs[Solved],sym_components);|
\]
Though for general nonlinear PDE systems, \verb|pdsolve| is known to sometimes return incomplete/non-general solution, this routine is one of the best ones in its class, and is rather efficient for linear determining equations. In particular, the dimension of the solution set (the number of free constants \verb|_Ci| in \verb|eq_tr_sol|) can be compared with \verb|simplified_eqs[dimension]| from Step F.

\bigskip\noindent \textbf{Step H. Output the equivalence transformation generators.} For this purpose, one uses the following \verb|GeM| routine, which outputs both the generators and the respective free constants and/or free functions in the general solution  \verb|eq_tr_sol|.
\[
\verb|gem_output_symm(eq_tr_sol);|
\]

\bigskip\noindent \textbf{Step I. Generate global group(s) of equivalence transformations.} In order to generate the global group of equivalence transformations corresponding to a given infinitesimal generator within the general solution  \verb|eq_tr_sol|, one ``extracts" the corresponding generator. For example, in order to extract the transformation corresponding to \verb|_C6|, one calls the routine
\[
\verb|Particular_Transform:=gem_extract_symm(sym_sol,spec={_C6=1});|
\]
(This routine sets free constants \verb|_Ci| within \verb|eq_tr_sol| not mentioned in \verb|spec| to zero.) Then the Lie theorem ODEs \eqref{eq:EQ:Lie1} are generated,
\[
\barr
\verb|Particular_Transform_ODEs:=gem_global_group(|\\
\verb|     Particular_Transform, group_param_name='a');|
\earr
\]
and solved:
\[
\verb|Particular_Transform_Global:=dsolve(Particular_Transform_ODEs);|
\]
The procedure of Step I can be repeated for the remaining linearly independent equivalence transformation generators. In some cases, in particular, when the local equivalence transformation group is not too large, it may be beneficial to skip the \verb|gem_extract_symm| step, and generate the global group ODEs and integrate them for the general solution \verb|eq_tr_sol| instead of the one given by \verb|Particular_Transform|\,.

\begin{remark}
We note that the order and/or the form of solutions obtained in Steps G, H, I may differ in different versions of \verb|Maple|, and also depending on the naming of variables a user employs. This has to do with sorting of variables and the introduction of arbitrary constants by \verb|pdsolve|. The final solution (i.e., the infinitesimal generators) will, in any case, provide an equivalent basis of the Lie algebra of infinitesimal generators for the sought generalized equivalence transformations.
\end{remark}

\begin{remark}
From the computational point of view, it is important to remark that in general, the generalized extended equivalence transformations do not form a Lie group of point transformations even in the space of variables extended by arbitrary elements. For example, the transformations defined by \eqref{eq:NLT:eg:Lisle:equiv:Tr}, \eqref{eq:NLT:eg:Lisle:equiv:Tr:BC} do not form a Lie group in the space of $(x, t, u, v, b, c)$, and hence do not belong to the set of generalized equivalence transformations \eqref{eq:ext:Melesh}.  It follows that the systematic procedure described in a current section will generally miss generalized extended equivalence transformations.
\end{remark}


\section{Examples of Symbolic Computation of Equivalence Transformations}\label{sec:eqtr:egs}

\subsection{Equivalence Transformations of the Dimensional KdV Family }\label{sec:eqtr:KdV}

We begin with a simple example of computing Lie groups of equivalence transformations of the dimensional KdV family \eqref{eq:dimKdV:fam}, according to the sequence of steps presented in Section \ref{sec:algo:gem}. The program code can have, for example, the following form.

Initialization: same as \eqref{eq:gem:initeq}.

Declaration of variables:
\[
\barr
\verb|ind:=x,t;|\\
\verb|dep:=U(ind),A(ind),B(ind),Q(ind);|\\
\verb|gem_decl_vars(indeps=[ind], deps=[dep]);|
\earr
\]
(capital letters are used for the parameters $a,b,q$ in \eqref{eq:dimKdV:fam}).

Declaration of the PDE family, solved for the leading derivative $u_t$:
\[
\barr
\verb|KdVfamily:=diff(U(ind),t)+A(ind)*diff(U(ind),x)|\\
\verb|    +B(ind)*U(x,t)*diff(U(ind),x)+Q(ind)*diff(U(ind),x,x,x)=0;|\\
\verb|gem_decl_eqs([KdVfamily], solve_for=[diff(U(ind),t)]);|
\earr
\]

Computation of the split system of equivalence transformation determining equations, using \verb|ind,dep| declared above, and output of infinitesimal components:
\[
\barr
\verb"det_eqs:=gem_symm_det_eqs([ind,dep]);"\\
\verb|OFF: sym_components:=gem_symm_components(); ON:|
\earr
\]
This step returns 182 determining equations stored in \verb|det_eqs|.

Restriction equations are obtained by demanding that the infinitesimal equivalence transformation components for the constitutive parameters $a, b, q$ may only depend on those parameters themselves.
\[
\barr
\verb|restriction_eqs:=gem_generate_EquivTr_dependence([|\\
\verb|    [[x,t],[x,t,U,A,B,Q]], |\\
\verb|    [[U],[x,t,U]],|\\
\verb|    [[A,B,Q],[A,B,Q]]|\\
\verb|]);|
\earr
\]

The full set of determining equations is created and simplified as follows:
\[
\barr
\verb|det_eqs_all:=det_eqs union restriction_eqs:|\\
\verb|simplified_eqs := DEtools[rifsimp](det_eqs_all, sym_components, |\\
\verb|     casesplit, mindim = 1);|
\earr
\]
This step results in 36 simplified linear PDEs (use \verb|linalg[vectdim](simplified_eqs[Solved]);|) on the equivalence transformation components. The dimension of the solution set of this system, i.e., the dimension of the Lie algebra of infinitesimal generators of equivalence transformations in the chosen form, is contained in the variable $\verb|simplified_eqs[dimension]|$ and equals 7.

The equivalence transformation components are solved for and displayed using the commands
\beq\label{mapl:solv:outp}
\barr
\verb|eq_tr_sol:=pdsolve( simplified_eqs[Solved],sym_components);|\\
\verb|gem_output_symm(eq_tr_sol);|
\earr
\eeq
which outputs seven generators corresponding, up to scaling constants, to \eqref{eq:1:sc2X} (sorted here according to the output expression length as a text string):
\[
\begin{array}{c}
\verb|_C6|,\quad  \sg{X}_1 = \dfrac{\partial }{\partial x},\\[2ex]
\verb|_C7|,\quad  \sg{X}_2 = \dfrac{\partial }{\partial t},\\[2ex]
\verb|_C2|,\quad  \sg{X}_3 = \dfrac{\partial }{\partial u} -b \dfrac{\partial }{\partial a} ,\\[2ex]
\verb|_C5|,\quad  \sg{X}_4 = t \dfrac{\partial }{\partial x} + \dfrac{\partial }{\partial a},\\[2ex]
\verb|_C3|,\quad  \sg{X}_5 = \dfrac{1}{2}\, t \,\dfrac{\partial }{\partial t} + \dfrac{1}{2}\,x\, \dfrac{\partial }{\partial x}  + q \,\dfrac{\partial }{\partial q}, \\[2ex]
\verb|_C1|,\quad  \sg{X}_6 = -\dfrac{3}{2}\,t\, \dfrac{\partial }{\partial t} - \dfrac{1}{2}\,x\,\dfrac{\partial }{\partial x}+ u\,\dfrac{\partial }{\partial u} +a \,\dfrac{\partial }{\partial a}, \\[2ex]
\verb|_C4|,\quad  \sg{X}_7 =  -\dfrac{3}{2}\,t\,\dfrac{\partial }{\partial t} - \dfrac{1}{2}\,x\, \dfrac{\partial }{\partial x} + a\,\dfrac{\partial }{\partial a}+b\,\dfrac{\partial }{\partial b}.
\end{array}
\]

In order to generate a global group of equivalence transformations corresponding, for example,  to the constant $A_7$ in \eqref{eq:KdV:eqivtr} (integration constant \verb|_C4| above), one uses the commands
\[
\barr
\verb|Particular_Transform:=gem_extract_symm(sym_sol,spec={_C4=1});|\\
\verb|Particular_Transform_ODEs:=gem_global_group(|\\
\verb|     Particular_Transform, group_param_name='s');|\\
\verb|Particular_Transform_Global:=dsolve(Particular_Transform_ODEs);|
\earr
\]
and obtains the global transformation
\beq\label{eq:dKdv:outp:globgroupA7}
\barr
\verb|A(s) = A_0*exp(s), B(s) = B_0*exp(s), Q(s) = Q_0, U(s) = U_0,|\\
\verb|t(s) = t_0*exp(-(3/2)*s), x(s) = x_0*exp(-(1/2)*s)|
\earr
\eeq
with the group parameter $s$, and the values $t_0$, $x_0$, etc., corresponding to the values of the variables before the transformation. The constant $A_7$ in \eqref{eq:KdV:eqivtr} is identified with $e^{s/2}$ in \eqref{eq:dKdv:outp:globgroupA7}, \verb|B_0| of \eqref{eq:dKdv:outp:globgroupA7} with $b$ in \eqref{eq:KdV:eqivtr}, \verb|B(s)| with $b^*$, etc. (For details, see \verb|E_g_01_equiv_tr_KdV_dim_ABQ.mw|.)

\subsection{Nonlocal Equivalence Transformations of a Family of Nonlinear Wave Equations}\label{sec:eqtr:NLWave}

Consider a family \eqref{eq:ch4:nonl_wave_ch4} of (1+1)-dimensional nonlinear wave equations for $u(x,t)$, involving an arbitrary constitutive function $c(u)$.
Local equivalence transformations of \eqref{eq:ch4:nonl_wave_ch4} are given by \eqref{eq:ch4:nonl_wave_equivt} with infinitesimal generators \eqref{eq:wav:ET:U}.

The PDE \eqref{eq:ch4:nonl_wave_ch4} admits  a local conservation law \eqref{eq:UW:CL}, and consequently a potential system \eqref{eq:ch4:nonl_wave_ch4:w}, equivalent but nonlocally related to the wave equation \eqref{eq:ch4:nonl_wave_ch4} for every $c(u)$ \cite{BC1,BC2}.

%

We now use the algorithm presented in Section \ref{sec:algo:gem} to compute the point equivalence transformations \eqref{eq:wav:ET:UW} of the potential system \eqref{eq:ch4:nonl_wave_ch4:w}.

After the initialization \eqref{eq:gem:initeq}, the program code is as follows. (Note that the extended list of ``dependent variables" for the computation includes the constitutive function $c(u)$ denoted by the capital $C$.)
\[
\barr
\verb|ind:=x,t;|\\
\verb|dep:=U(ind),W(ind),C(ind);|\\
\verb|gem_decl_vars(indeps=[ind], deps=[dep]);|
\earr
\]
The equations are declared as follows.
\[
\barr
\verb|UW1:=diff(W(ind),x)=t*diff(U(ind),t) - U(ind);|\\
\verb|UW2:=diff(W(ind),t)=t*C(ind)^2*diff(U(ind),x);|\\
\verb|gem_decl_eqs([UW1,UW2], solve_for=[lhs(UW1), lhs(UW2)]);|
\earr
\]
One may seek Lie groups of generalized equivalence transformations by requiring that the component for the constitutive function $c(u)$ is independent of $x,t,w$.
\[
\barr
\verb"det_eqs:=gem_symm_det_eqs([ind,dep]);"\\
\verb|OFF: sym_components:=gem_symm_components(); ON:|\\
\verb|restriction_eqs:=gem_generate_EquivTr_dependence([|\\
\verb|    [[x,t,U,W],[x,t,U,W,C]],|\\
\verb|    [[C],[U,C]] ]);|\\
\verb|det_eqs_all:=det_eqs union restriction_eqs:|\\
\verb|simplified_eqs := DEtools[rifsimp](det_eqs_all, sym_components, |\\
\verb|     casesplit, mindim = 1);|
\earr
\]
This leads to 25 simplified linear determining equations, which are subsequently solved to obtain an eight-dimensional Lie algebra of the equivalence transformation infinitesimal generators.
\[
\barr
\verb|eq_tr_sol:=pdsolve( simplified_eqs[Solved],sym_components);|\\
\verb|gem_output_symm(eq_tr_sol);|
\earr
\]
The output (in \verb|Maple| 18; may slightly vary across versions) is
\beq\label{eq:wav:ET:UW:symb}
\begin{array}{c}
\verb|_C3|,\quad  \sg{X}_1 =  \dfrac{\partial }{\partial w},\\[2ex]
\verb|_C8|,\quad  \sg{X}_2 =  \dfrac{\partial }{\partial x},\\[2ex]
\verb|_C4|,\quad  \sg{X}_3 = t \,\dfrac{\partial }{\partial u},\\[2ex]
\verb|_C1|,\quad  \sg{X}_4 = x \dfrac{\partial }{\partial w} - \dfrac{\partial }{\partial u},\\[2ex]
\verb|_C7|,\quad  \sg{X}_5 = t \,\dfrac{\partial }{\partial t}-c\,\dfrac{\partial }{\partial c},\\[2ex]
\verb|_C2|,\quad  \sg{X}_6 = t\,\dfrac{\partial }{\partial t} + x\,\dfrac{\partial }{\partial x} + w\, \dfrac{\partial }{\partial w}, \\[2ex]
\verb|_C5|,\quad  \sg{X}_7 = u\,\dfrac{\partial }{\partial u}- t\,\dfrac{\partial }{\partial t} - x\,\dfrac{\partial }{\partial x}, \\[2ex]
\verb|_C6|,\quad  \sg{X}_8 = -2cu\,\dfrac{\partial }{\partial c} + u^2\,\dfrac{\partial }{\partial u} +tu\,\dfrac{\partial }{\partial t} + w\,\dfrac{\partial }{\partial x},
\end{array}
\eeq
which (up to renumbering and rescaling) coincides with the generators listed in the formulas \eqref{eq:wav:ET:UW}.

As an illustration, we now compute a global group of equivalence transformations corresponding to the projective transformation $\sg{W}_8$ of \eqref{eq:wav:ET:UW}, which corresponds to  the integration constant \verb|_C6| in the output \eqref{eq:wav:ET:UW:symb}. One proceeds as follows.
\[
\barr
\verb|Projective_Transf:=gem_extract_symm(sym_sol,spec={_C6=-1});|\\
\verb|Particular_Transform_ODEs:=gem_global_group(|\\
\verb|     Projective_Transf, group_param_name='B');|\\
\verb|Particular_Transform_Global:=dsolve(Particular_Transform_ODEs);|
\earr
\]
and obtains the output
\[
\barr
\verb|x(B) = x_0-B*W_0, t(B)=t_0/(1+BU_0), U(B) = U_0/(1+BU_0),|\\
\verb|w(B)=W_0, c(B)=c_0*(1+BU_0)^2|
\earr
\]
corresponding to the global transformation \eqref{eq:ch4:NLW:UW_equivt2}
with the group parameter $B$. \\(See \verb|E_g_02_equiv_tr_waveUW.mw|.)

\subsection{Generalized Equivalence Transformations of the KdV-Burgers Family }\label{sec:eg:BKdV}

Consider a family \eqref{eq:KdVB} of (1+1)-dimensional KdV-Burgers equations involving $r+3$ arbitrary elements that are functions of $x,t$. In \cite{Opanasenko}, equivalence group computations were done through tedious algebraic manipulation. For the current computational example, following \cite{Opanasenko}, we fix $r=3$ and choose $A_1=0$, $A_3=1$. The explicit PDE family is then given by
\beq\label{eq:KdVB:eg}
u_t + C u u_x = A_0 u + A_2 u_{xx} + u_{xxx} + B,
\eeq
with four arbitrary elements  $A_0(x,t)$, $A_2(x,t)$, $B(x,t)$, $C(x,t)$. First we compute the Lie group of point generalized equivalence transformations of \eqref{eq:KdVB:eg}.

\medskip\noindent\textbf{a) Point equivalence transformations.}  After the initialization \eqref{eq:gem:initeq}, the program code is as follows. The variables and the PDE family are declared:
\beq\label{eq:KdVB:eg:inddep1}
\barr
\verb|ind:=x,t;|\\
\verb|dep:=U(ind),A0(ind),A2(ind),B(ind),C(ind);|\\
\verb|gem_decl_vars(indeps=[ind], deps=[dep]);|
\earr
\eeq

\beq\label{eq:KdVB:eg:pdes1}
\barr
\verb|BKdV:=diff(U(ind),t)+C(ind)*U(ind)*diff(U(ind),x)=A0(ind)*U(ind)|\\
\verb|  + A2(ind)*diff(U(ind),x,x)+diff(U(ind),x,x,x) +B(ind);|\\
\verb|gem_decl_eqs([BKdV], solve_for=[diff(U(ind),t)]);|
\earr
\eeq
We seek most general Lie groups of point generalized equivalence transformations with generators in the form
\beq\label{eq:EQ:X}
\sg{X} = \xi^x\frac{\partial }{\partial x} + \xi^t\frac{\partial }{\partial t}+\eta^u \frac{\partial }{\partial u}+\theta^{A_0}\frac{\partial }{\partial A_0}+\theta^{A_2}\frac{\partial }{\partial A_2} +\theta^B\frac{\partial }{\partial B} +\theta^C\frac{\partial }{\partial C},
\eeq
where the components $\theta^j$ corresponding to the arbitrary functions $A_0(x,t)$, $A_2(x,t)$, $B(x,t)$, $C(x,t)$ may depend on $x$, $t$, $A_0$, $A_2$, $B$, $C$, and the components $\xi^t$, $\xi^x$ and $\eta^u$ corresponding to the variables may depend on $x$, $t$, $u$, $A_0$, $A_2$, $B$, $C$. The restriction equations become
\beq\label{eq:KdVB:eg:solv1}
\barr
\verb|restriction_eqs:=gem_generate_EquivTr_dependence([|\\
\verb|    [[C,A0,A2,B],[x,t,A0,A2,B,C]] ]);|\\
\verb|det_eqs_all:=det_eqs union restriction_eqs:|\\
\earr
\eeq
\beq\label{eq:KdVB:eg:solv2}
\barr
\verb|simplified_eqs := DEtools[rifsimp](det_eqs_all, sym_components, |\\
\verb|     casesplit, mindim = 1);|
\earr
\eeq
This leads to 36 simplified linear determining equations, which are subsequently solved and output using commands \eqref{mapl:solv:outp}. The output (in \verb|Maple| 18; may slightly vary across versions) is
\beq\label{eq:BKdV:symb:geq:1}
\begin{array}{c}
\verb|_C2|,\quad  \sg{X}_1 =  \dfrac{\partial }{\partial x},\\[2ex]
\verb|_C3|,\quad  \sg{X}_2 =  \dfrac{\partial }{\partial t},\\[2ex]
\verb|_F1|,\quad  \sg{X}_3 = F_1'(t) \,\dfrac{\partial }{\partial A_0}+ F_1(t) \left( u \dfrac{\partial }{\partial u} + B \dfrac{\partial }{\partial B} - C \dfrac{\partial }{\partial C}\right),\\[2ex]
\verb|_C1|,\quad  \sg{X}_4 = \dfrac{3}{2}\,\left( -t \dfrac{\partial }{\partial t} + A_0\dfrac{\partial }{\partial A_0}+ B\dfrac{\partial }{\partial B}\right)+\dfrac{1}{2}\,\left( -x \dfrac{\partial }{\partial x} + A_2\dfrac{\partial }{\partial A_2}\right).\\[2ex]
\end{array}
\eeq
Here in addition to scalings ($\sg{X}_4$) one has an equivalence transformation involving an arbitrary  function of $t$, but no generalized equivalence transformations \eqref{eq:KdVB:eqivtr}, \eqref{eq:KdVB:eqivtr2} found in \cite{Opanasenko}. For details, see an accompanying Maple file \verb|E_g_03a_BKdV_point_eq_tr.mw|

\medskip\noindent\textbf{b) Generalized equivalence transformations.}  In order to compute the generalized transformations \eqref{eq:KdVB:eqivtr}, \eqref{eq:KdVB:eqivtr2}, one must allow the infinitesimal transformation components for the arbitrary element $B(x,t)$ to depend on $C_t$, $C_x$, $C_{xx}$, and $C_{xxx}$ \cite{Opanasenko}. The above code in \eqref{eq:KdVB:eg:inddep1}, \eqref{eq:KdVB:eg:pdes1}, and \eqref{eq:KdVB:eg:solv1} is modified as follows. First, additional dependent variables $C_{t}$, $\ldots$, $C_{xxx}$ are declared:
\[
\barr
\verb|dep:=U(ind),A0(ind),A2(ind),B(ind),C(ind),|\\
\verb|CT(ind),CX(ind),CXX(ind),CXXX(ind);|
\earr
\]
Differential relationships between $C$, $C_{t}$, $\ldots$, $C_{xxx}$ are then added to the PDE system:
\[
\barr
\verb|gem_decl_eqs([BKdV,  diff(C(ind),t)=CT(ind),  diff(C(ind),x)=CX(ind),|\\
\verb|   diff(CX(ind),x)=CXX(ind),  diff(CXX(ind),x)=CXXX(ind)],|\\
\verb|solve_for=[diff(U(ind),t),  diff(C(ind),t),  diff(C(ind),x),|\\
\verb|   diff(CX(ind),x),  diff(CXX(ind),x)]);|
\earr
\]
Restriction equations are modified:
\[
\barr
\verb|restriction_eqs:=gem_generate_EquivTr_dependence([|\\
\verb|    [[x,t],[x,t]],|\\
\verb|    [[U],[x,t,U,C]],|\\
\verb|    [[A0,A2,B,C,CT,CX,CXX,CXXX],[x,t,A0,A2,B,C,CT,CX,CXX,CXXX]] ]);|\\
\earr
\]
Then the simplification and solution of the overdetermined system \eqref{eq:KdVB:eg:solv2} takes place. The output includes and extends the generators given in \eqref{eq:BKdV:symb:geq:1}. In particular, it includes an additional generalized equivalence transformation generator \eqref{eq:EQ:X} with
\[
\eta^u = \eta^u(x,t,u,C) = F_6(t) u + \dfrac{F_3'(t)x +F_7'(t)}{C},
\]
featuring an explicit dependence on the arbitrary element $C$ (cf. \eqref{eq:KdVB:eqivtr}, \eqref{eq:KdVB:eqivtr2}). For details, see an accompanying Maple file \verb|E_g_03b_BKdV_generalized_eq_tr.mw|.



%
%


\section{Equivalence Transformations of a Family of Nonlinear Equations Describing Anti-Plane Shear Motions of a Fiber-Reinforced Material}\label{sec:eqtr:1fiber}

The framework of incompressible hyper/viscoelasticity, in particular, its formulations that take into account possible anisotropy due to the presence of a fiber matrix, provide a practically important toolbox  for the description of a various classes of materials, including applications in medicine (biological membranes), cell biology (cell membranes), industry (textiles and ribbers), and more \cite{cheviakov2015one}. Partial differential equations describing finite deformations of such solids are force balance-type (3+1)-dimensional equations for the time evolution of Eulerian positions of Largangian points of the solid. The model is formulated through the definition of invariants and pseudo-invariants that may involve deformation tensors, deformation rates, and fiber directors. In the incompressible setting, in three dimensions, the model is given by four PDEs (three equations of motion and the incompressibility condition). The arbitrary elements present in such models are contained in the form of the strain energy density function $W$, and are given by contributions of one or more arbitrary functions describing various aspects of the material behaviour. For example, the general form of the strain energy density in the incompressible hyperelastic setting is given by
\[
W=W_{iso}(I_1,I_2) + W_{aniso}(I_4,\ldots, I_9),
\]
where $W_{iso}$ and $W_{aniso}$ are isotropic and anisotropic contributions, respectively; $I_1,I_2$ are invariants of the left and right Cauchy–Green deformation tensors (the third invariant $I_3=J^2=1$, the unit Jacobian of the incompressible deformation tensor), and $I_4,\ldots, I_9$ are pseudo-invariants involving fiber direction vectors (for details, see, e.g., \cite{cheviakov2015fully,cheviakov2015one}, and references therein).

A basic yet important set of physical systems describing finite displacements of fiber-reinforced Mooney-Rivlin materials under the standard reinforcing model \cite{cheviakov2015fully} is formulated in terms of the strain energy density
\beq\label{eq:anz:1fib:hyperel:gengam:W}
W=a(I_1-3)+b(I_2-3)+q(I_4 - 1)^{2},
\eeq
where $a,\,b,\,q>0$ are constant material parameters.

In \cite{cheviakov2015fully}, a family of nonlinear wave equations has been derived, describing one-dimensional S-wave propagation dependent on the fiber direction, within the above framework:
\beq\label{eq:anz:1fib:hyperel:gengam:PDE}
G_{tt} = \left( \alpha
	+ {\beta\cos^{2}\gamma  \left( 3\cos^{2}\gamma  \left(G_x\right)^{2} + 6\sin\gamma \cos\gamma  \,G_x + 2\sin^2\gamma  \right) }\right)  G_{xx}.
\eeq
Here $G(x,t)$ describes the finite displacement amplitude of anti-plane shear motions, in the material $z$-direction, of a nonlinear incompressible hyperelasic medium, reinforced by fibers located in the $(x,z)$-plane of the material frame. The fibers are assumed to make a constant angle $\gamma$ with the material direction $x$. The model involves three arbitrary elements, the material parameters $\alpha=2(a+b)>0$,  $\beta=4q>0$, and the material fiber angle $\gamma\in[0,\pi/2]$.

When $\gamma=\pm \pi/2$, i.e., the fibers are aligned along the material $z$-direction, and the PDEs \eqref{eq:anz:1fib:hyperel:gengam:PDE} reduce to the classical linear wave equations $G_{tt} = \alpha G_{xx}$, in particular, fiber effects vanish. When the fiber angle is $\gamma=0$, equations \eqref{eq:anz:1fib:hyperel:gengam:PDE} assume a simple form
\beq\label{eq:anz:1fib:hyperel:PDE:gam0}
G_t = (\alpha+3\beta G_x^2)\,G_{xx}.
\eeq


We now use the algorithm of Sections \ref{sec:algo:equi} and \ref{sec:algo:gem} to compute equivalence  transformations of the PDE family \eqref{eq:anz:1fib:hyperel:gengam:PDE}. The PDEs involve three arbitrary elements; we assume $\gamma\ne \pm \pi/2$. For the sake of simplicity in \verb|Maple| notation, we denote $\alpha=a$, $\beta=b$; moreover, in order to simplify computations, as we will see, it is beneficial to use the arbitrary element $k=\tan\gamma\in \mathbb{R}$ instead of $\gamma$ itself. The corresponding infinitesimal generators of point equivalence transformations consequently assume the form
\beq \label{eq:AntiPlaneeq_trans:X}
\sg{X} =\xi \dfrac{\partial}{\partial x}+\tau \dfrac{\partial}{\partial t}+\eta \dfrac{\partial}{\partial G} +\zeta^1 \dfrac{\partial}{\partial a}+\zeta^2\dfrac{\partial}{\partial b}+\zeta^3 \dfrac{\partial}{\partial k}.
\eeq
For the current application, it is sufficient to consider regular equivalence transformations \eqref{eq:ch1:symm:equivalence:eq_trans}, so the dependencies of the infinitesimal components are chosen as follows:
\[
\xi =\xi (x,t,G),\qquad \tau =\tau (x,t,G),\qquad \eta =\eta (x,t,G),
\]
\[
\zeta^\nu =\zeta^\nu (a,b,k),\quad \nu=1,2,3.
\]

The program sequence for equivalence transformation computations proceeds as follows. The \verb|Maple| modules are initialized according to \eqref{eq:gem:initeq}. The PDE family \eqref{eq:anz:1fib:hyperel:gengam:PDE}, independent variables, the dependent variable, and arbitrary elements are defined as variables of the problem.
\[
\barr
\verb|PDE1:=diff(G(x,t), t, t) = (a + b*cos(g)^2* |\\
\verb|    ( 3*cos(g)^2*diff(G(x,t), x)^2 +6*sin(g)*cos(g)*diff(G(x,t), x)|\\
\verb|      + 2*sin(g)^2))* diff(G(x,t), x, x);|
\earr
\]
\[
\barr
\verb|ind:=x,t;|\\
\verb|a:=A(ind); b:=B(ind); k:=K(ind);|\\
\verb|dep:=G(ind), a,b,k;|\\
\earr
\]

The \verb|rifsimp| routine that will be used for the determining equation reduction, by design, can only process differential polynomials, yet the sine and cosine functions present in the PDEs \eqref{eq:anz:1fib:hyperel:gengam:PDE} are not polynomial functions of the constitutive parameter $k=\tan\gamma$. This restriction can be overcome by temporarily treating the  sine and cosine as ``arbitrary functions":
\[
\barr
\verb|subs_sin_cos := {cos(g) = CG(K(x, t)), sin(g) = SG(K(x, t))};|\\
\verb|PDE2:=subs(subs_sin_cos,PDE1);|\\
\earr
\]
Finally, the variables of the problem and the PDE family is declared:
\[
\barr
\verb|gem_decl_vars(indeps=[ind], deps=[dep]);|\\
\verb|gem_decl_eqs([PDE2], solve_for=[diff(G(x,t),t,t)]);|
\earr
\]

As usual, we seek generalized equivalence transformations, initially allowing all infinitesimal components to depend on all variables of the problem, and display these infinitesimal component dependencies.
\[
\barr
\verb|det_eqs:=gem_symm_det_eqs([ind, dep]):|\\
\verb|OFF: sym_components:=gem_symm_components(); ON:|\\
\earr
\]
This yields 187 determining equations. To seek usual equivalence transformations, we restrict the dependencies of local transformation components as follows:
\[
\barr
\verb|restriction_eqs:=gem_generate_EquivTr_dependence([|\\
\verb|    [[x,t],[x,t]],|\\
\verb|    [[G],[x,t,G,A,B,K]],|\\
\verb|    [[A, B, K],[A, B, K]]]);|\\
\earr
\]
Moreover, a relationship ensuring the proper treatment of $\sin\gamma$, $\cos\gamma$ in relation to $k=\tan\gamma$ is imposed:
\[
\barr
\verb|cos_sin_cond:={CG(K)^2=1/(K^2+1),SG(K)^2=K^2/(K^2+1)};|
\earr
\]
Importantly, the relationships \verb|cos_sin_cond| are equivalent to polynomial ones, and can be handled by \verb|rifsimp|. Finally, the equivalence transformation determining equations are given by the set
\[
\barr
\verb|det_eqs1:=det_eqs union restriction_eqs union cos_sin_cond:|
\earr
\]
This set of equations is simplified using the command
\[
\barr
\verb|simplified_eqs := DEtools[rifsimp](det_eqs1, sym_components, |\\
\verb|     casesplit, mindim = 1);|
\earr
\]
which outputs 39 simplified PDEs, and computes the dimension of the solution space, i.e., the number of equivalence transformations in the specified ansatz: \verb|simplified_eqs[dimension]=8|.

Finally, the simplified system of determining equations remains to be solved. Before that, we substitute the actual values for the sine and cosine functions, keeping in mind that $\gamma\in[0,\pi/2]$.
\[
\barr
\verb|solve_sin_cos:=allvalues(solve(cos_sin_cond,{CG(K),SG(K)}));|\\
\verb|simplified_eqs2 := eval(subs(solve_sin_cos[1], simplified_eqs[Solved])):|
\earr
\]
The solution and output of the equivalence transformation generators proceeds as follows:
\[
\barr
\verb|eq_tr_sol:=pdsolve(simplified_eqs2, sym_components);|
\verb|gem_output_symm(eq_tr_sol);|
\earr
\]
The output (in \verb|Maple| 18) is
\beq\label{eq:wav:ET:UW:symb}
\begin{array}{c}
\verb|_C5|,\quad  \sg{X}_1 = \dfrac{\partial }{\partial G},\\[2ex]
\verb|_C7|,\quad  \sg{X}_2 = \dfrac{\partial }{\partial x},\\[2ex]
\verb|_C8|,\quad  \sg{X}_3 = \dfrac{\partial }{\partial t},\\[2ex]
\verb|_C3|,\quad  \sg{X}_4 = t\,\dfrac{\partial }{\partial G},\\[2ex]
\verb|_C4|,\quad  \sg{X}_5 = t\,\dfrac{\partial }{\partial t}+x\,\dfrac{\partial }{\partial x}+G\,\dfrac{\partial }{\partial G},\\[2ex]
\verb|_C6|,\quad  \sg{X}_6 = -\dfrac{1}{2}\,t\,\dfrac{\partial }{\partial t}+a\,\dfrac{\partial }{\partial a}+b\,\dfrac{\partial }{\partial b},\\[2ex]
\verb|_C1|,\quad  \sg{X}_7 = -x\,\dfrac{\partial }{\partial x} -2 a\,\dfrac{\partial }{\partial a} - \dfrac{4b}{k^2+1}\,\dfrac{\partial }{\partial b}+k\,\dfrac{\partial }{\partial k}, \\[2ex]
\verb|_C2|,\quad  \sg{X}_8 = -x\,\dfrac{\partial }{\partial G} + \dfrac{2bk}{(k^2+1)^2}\,\dfrac{\partial }{\partial a} + \dfrac{4bk}{k^2+1}\,\dfrac{\partial }{\partial b}+\dfrac{\partial }{\partial k}
\end{array}
\eeq
(for details, see \verb|E_g_04_equiv_tr_elast_antiplane_tan.mw|).
The transformation generator $\sg{X}_8$ in \eqref{eq:wav:ET:UW:symb} corresponds to shifts in $k\equiv\tan\gamma$. The global group is found in \verb|Maple| using the commands
\[
\barr
\verb|Particular_Transform:=gem_extract_symm(sym_sol,spec={_C8=1});|\\
\verb|Particular_Transform_ODEs:=gem_global_group(|\\
\verb|     Particular_Transform, group_param_name='S');|\\
\verb|Particular_Transform_Global:=dsolve(Particular_Transform_ODEs);|
\earr
\]
and is given by
\[
\barr
{G}^* = G - Sx, \qquad \tan{\gamma}^* = \tan \gamma + S,\\[2ex]
{\alpha}^* =  \alpha + 2\beta\cos^{4}\gamma\, \Big(\dfrac{s^2}{2}+S\tan\gamma\Big),\qquad
{\beta}^* =  \beta\cos^{4}\gamma\, \Big(\tan^2\gamma +2S\tan\gamma +S^2+1\Big)^2,
\earr
\]
with a group parameter $S\in \mathbb{R}$. In particular, the choice of $S=-k=-\tan \gamma$ lets one invertibly map the PDE \eqref{eq:anz:1fib:hyperel:gengam:PDE} into the form \eqref{eq:anz:1fib:hyperel:PDE:gam0} with $\gamma^*=0$. This equivalence transformation is given by
\beq\label{eq:Tr1DtoG0}
x= {x}^*,\qquad t= {t}^*,\qquad G={G}^* -x\tan\gamma,\qquad \alpha = {\alpha}^* + {\beta}^*\tan^2\gamma, \qquad \beta= {\beta}^* \cos^{-4}\gamma.
\eeq
The substitution of \eqref{eq:Tr1DtoG0} into the full PDE \eqref{eq:anz:1fib:hyperel:gengam:PDE} yields
\beq\label{eq:anz:1fib:hyperel:gam0:PDE:tildes}
{G}^*_{t^*t^*} = ({\alpha}^* + 3{\beta}^* ({G}^*_{{x^*}})^2) \,{G}^*_{x^*x^*}
\eeq
with
\[
{\alpha}^*= \alpha + \dfrac{\beta}{4}\sin^2 2\gamma,\qquad {\beta}^* = \beta \cos^{4}\gamma.
\]
which is the equation \eqref{eq:anz:1fib:hyperel:PDE:gam0} up to the omission of the tildes.

We note that the equation \eqref{eq:anz:1fib:hyperel:gam0:PDE:tildes} involves the parameter ${\alpha}^*$ that is not necessarily positive. In particular, loss of hyperbolicity may occur when the coefficient of $G_{xx}$ becomes negative \cite{cheviakov2015one}. A sufficient condition for the loss of hyperbolicity not to happen under the indicated transformation is given by
\beq\label{eq:cond:NoHypLoss}
\sin^2(2\gamma)< \dfrac{4\alpha}{\beta}.
\eeq
The condition \eqref{eq:cond:NoHypLoss} is satisfied for all fiber orientations $\gamma$, for example, in the common situation of relatively weak fiber contributions, ${\beta}<{4\alpha}$. In this case, obvious non-dimensionalizing scaling transformations
\beq\label{eq:fiber:rescale}
\widehat{x}=\dfrac{x^*}{L},\qquad \widehat{t}=\dfrac{\sqrt{\alpha^*}t^*}{L},\qquad\widehat{G}=\sqrt{\dfrac{3\beta^*}{\alpha^*}}\dfrac{G^*}{L},\qquad L\in \mathbb{R}^{+}
\eeq
of the PDE family \eqref{eq:anz:1fib:hyperel:gam0:PDE:tildes} can be used to further map these equations into a single nonlinear wave equation
\beq\label{eq:anz:1fib:hyperel:no:const}
\widehat{G}_{\widehat{t}\widehat{t}} = (1+(\widehat{G}_{\widehat{x}})^2) \,\widehat{G}_{\widehat{x}\widehat{x}}.
\eeq
Thus through systematically computed equivalence transformations \eqref{eq:Tr1DtoG0} and \eqref{eq:fiber:rescale}, the PDE family \eqref{eq:anz:1fib:hyperel:gengam:PDE} is invertibly reduced to a wave equation \eqref{eq:anz:1fib:hyperel:no:const} that involves no arbitrary elements. Properties of the PDE \eqref{eq:anz:1fib:hyperel:no:const} and the extended versions of the model are discussed in \cite{cheviakov2015one}.

\section{Discussion}\label{sec:concl}

The current contribution describes a practical algorithm of computation of Lie groups of point equivalence transformations and generalized equivalence transformations of families of differential equations, and a symbolic implementation of that algorithm in the \verb|GeM| package for \verb|Maple|. The method is directly applicable to DE families with arbitrary elements in the form of constants or functions of independent and dependent variables. As discussed in Section \ref{sec:equiv:theor}, the computation is based on a replacement of arbitrary elements by effective extra dependent variables, and a computation of Lie groups of point symmetries of such a modified DE system, with appropriate additional conditions imposed on the transformation  components corresponding to the arbitrary elements. The steps for symbolic computations are outlined in Section \ref{sec:algo:gem}. In particular, the simplification of an overdetermined system of equivalence transformation determining equations is done using the highly efficient and robust \verb|rifsimp| routine built into \verb|Maple| software.

Section \ref{sec:eqtr:egs} contains basic examples of equivalence transformation computations using \verb|Maple|/\verb|GeM|. Equivalence transformations for the classical family of dimensional KdV equations, and nonlocal equivalence transformations of a general class of nonlinear wave equations were computed therein. Additionally, a first computational example of genuine generalized equivalence transformations was presented (Section \ref{sec:eg:BKdV}).

In Section \ref{sec:eqtr:1fiber}, equivalence transformations of a family of nonlinear equations describing anti-plane shear motions of a fiber-reinforced material were considered. For the first and the third example, it was shown how the equivalence transformations can be used to invertibly map a given DE family into a single equation of a significantly simpler form, involving no arbitrary elements. In order to simplify the presentation and interpretation, the examples chosen for the current manuscript were limited to a situation with one time and one spatial variable. However, the computation algorithm can be applied to ODEs as well as PDEs and systems thereof in multi-dimensions; the presented software can be used without any changes.

It is worth mentioning the computational efficiency of the algorithm within the \verb|GeM| package that generates the determining equations for local symmetries and equivalence transformations. For a usual PDE system arising in physics, involving one to four independent variables and a comparable number of dependent variables, a typical computation and splitting of symmetry/equivalence transformation determining equations, in a problem that does not involve classifications, is finished usually within seconds, or a fraction of a second, on a standard desktop computer. The numbers of split determining equations can substantially differ, depending on the DE system at hand. Examples in Sections \ref{sec:eqtr:egs} and \ref{sec:eqtr:1fiber} involve less than two hundred determining equations; similar numbers of determining equations arise, for example, in the study of point symmetries of magnetohydrodynamics equilibrium equations \cite{cheviakov2004bogoyavlenskij}. We note that both \verb|GeM| software and the \verb|Maple| \verb|rifsimp| routine can efficiently processes substantially more complicated cases; for example, one calculations involving 31,918 and 58,273 determining equations were successfully completed in a related application of local conservation law computations in fluid dynamics models \cite{cheviakov2014generalized}.

It is straightforward to extend the algorithm and the program sequence described in Section \ref{sec:algo:gem} to address equivalence transformation classification problems. An example of that kind would be a problem of classification of generalized equivalence transformations of the PDE family \eqref{eq:anz:1fib:hyperel:gengam:PDE}, with arbitrary elements $\alpha$ and $\beta$, with respect to the classifying parameter $\gamma$; in particular, for special values of $\gamma$, additional equivalence transformations may arise. The \verb|casesplit| option of the \verb|rifsimp| routine enables one to perform such classifications. Specific examples can be found in \cite{BCABook} (Section 5.4.2), with \verb|Maple| code examples available from \cite{GemReferenceOnline}.

In conclusion, we note that there exist a significant number of challenging questions both in the theory of equivalence transformations and in the realm of practical computation. We briefly outline some of them here.
\begin{enumerate}
  \item Equivalence transformations of a family of DEs, understood in the general geometrical sense, are evidently not exhausted by Lie groups of point equivalence transformations or their generalizations discussed in Section \ref{subsec:eq:tr:gen}. The definition of broader classes of equivalence transformations which can be systematically analyzed remains an open problem, as does the analysis of completeness of equivalence transformation classifications within those classes.



  \item  A basic extension of the notion of point equivalence transformations is the notion of generalized equivalence transformations (Section \ref{subsec:eq:tr:gen}(A)). Examples where the local transformation components for dependent and independent variables involve constant arbitrary elements are well-known. To the best of authors's knowledge, the first example in the literature of generalized equivalence transformations where the transformations of dependent and/or independent variables involve arbitrary \emph{function(s)} of the model appears in \cite{Opanasenko}. The corresponding symbolic computations were performed in the current contribution in Section \ref{sec:eg:BKdV}.



  \item The so-called generalized extended equivalence transformations (see Section \ref{subsec:eq:tr:gen}B) provide significant, practically useful extensions of sets of equivalence transformations admitted by certain families of differential equations. It is necessary to come up with systematic approaches to the computation of generalized extended equivalence transformations, and symbolic software-assisted calculation algorithms.



  \item Symbolic computations of symmetries and equivalence transformations of families of \emph{linear} partial differential equations present a substantial challenge, due to the presence of infinite-parameter symmetry groups. The same is true for DE families that admit linearlizations by point transformations. A review of theoretical results pertaining to the computation of symmetries of linear PDEs and their software implementation can be found in \cite{cheviakov2010symbolic}. In future work, it is important to develop a better understanding of the structure of Lie groups of (generalized) point equivalence transformations of linear PDE families, and extend software implementations for point symmetry computation of linear DEs to include the computation of some classes of equivalence transformations of families of linear equations.


  \item In some classes of PDE models that arise, for example, in models of nonlinear compressible elastodynamics, constitutive functions may depend on combinations of derivatives of dependent variables of the problem, or more generally, be differential functions of the dependent variables \cite{marsden1994mathematical, cheviakov2012symmetry}. The algorithm described in the current contribution can be used to compute generalized point equivalence transformations for such models where the transformations for the arbitrary elements depend only on those arbitrary elements themselves. It remains an open problem to find examples, general formulas, and an overall algorithmic approach to the computation of some broader classes of equivalence transformations for such problems, in particular, equivalence transformations with a general action on such arbitrary elements with complicated dependencies.

\end{enumerate}

\subsubsection*{Acknowledgements}

The author is grateful to NSERC of Canada for research support through a Discovery grant.

\bibliography{bibequiv_18}
\bibliographystyle{phcpc}

\end{document}